\documentstyle[epsf,12pt]{article}

\begin{document}
\bibliographystyle{unsrt}
\hfuzz=12pt
 
\font\twelvemb=cmmib10 scaled \magstep1
\font\tenmb=cmmib10
\font\ninemb=cmmib9
\font\sevenmb=cmmib7
\font\sixmb=cmmib6
\font\fivemb=cmmib5
\font\twelvesyb=cmbsy10 scaled \magstep1
\font\tensyb=cmbsy10
\font\sstwelve=cmss12
\font\ssnine=cmss9
\font\sseight=cmss8
\textfont9=\twelvemb
\scriptfont9=\tenmb
\scriptscriptfont9=\sevenmb
\textfont10=\twelvesyb
\def\bm{\fam9}
\def\bms{\fam10}

\newcommand{\DS}{\displaystyle}
\newcommand{\TS}{\textstyle}
\newcommand{\SS}{\scriptstyle}
\newcommand{\SSS}{\scriptscriptstyle}

\newcommand\zZtwelve{\hbox{\sstwelve Z\hskip -4.5pt Z}}
\newcommand\zZnine{\hbox{\ssnine Z\hskip -3.9pt Z}}
\newcommand\zZeight{\hbox{\sseight Z\hskip -3.7pt Z}}
\newcommand\zZ{\mathchoice{\zZten}{\zZten}{\zZeight}{\zZeight}}
\newcommand\ZZ{\mathchoice{\zZtwelve}{\zZtwelve}{\zZnine}{\zZeight}}

\mathchardef\sigma="711B
\mathchardef\tau="711C
\mathchardef\omega="7121
\mathchardef\nabla="7272

\newcommand{\e}{\epsilon}
\newcommand{\eps}{\epsilon}
\newcommand{\ee}{\varepsilon}
\newcommand{\vp}{\varphi}
\newcommand{\vphi}{\varphi}
\newcommand{\cphi}{\Phi}
\newcommand{\om}{\omega}
\let\oldvrho=\varrho
\newcommand{\vrho}{{\raise 2pt\hbox{$\oldvrho$}}}
\let\oldchi=\chi
\renewcommand{\chi}{{\raise 2pt\hbox{$\oldchi$}}}
\let\oldxi=\xi
\renewcommand{\xi}{{\raise 2pt\hbox{$\oldxi$}}}
\let\oldzeta=\zeta
\renewcommand{\zeta}{{\raise 2pt\hbox{$\oldzeta$}}}

\newcommand{\la}[1]{\label{#1}}
\newcommand{\ur}[1]{(\ref{#1})}
\newcommand{\ra}[1]{(\ref{#1})}
\newcommand{\urs}[2]{(\ref{#1},~\ref{#2})}
\newcommand{\eq}[1]{eq.~(\ref{#1})}
\newcommand{\eqs}[2]{eqs.~(\ref{#1},~\ref{#2})}
\newcommand{\eqss}[3]{eqs.~(\ref{#1},~\ref{#2},~\ref{#3})}
\newcommand{\eqsss}[2]{eqs.~(\ref{#1}--\ref{#2})}
\newcommand{\Eq}[1]{Eq.~(\ref{#1})}
\newcommand{\Eqs}[2]{Eqs.~(\ref{#1},~\ref{#2})}
\newcommand{\Eqss}[3]{Eqs.~(\ref{#1},~\ref{#2},~\ref{#3})}
\newcommand{\Eqsss}[2]{Eqs.~(\ref{#1}--\ref{#2})}
\newcommand{\fig}[1]{Fig.~\ref{#1}}
\newcommand{\figs}[2]{Figs.~\ref{#1},\ref{#2}}
\newcommand{\figss}[3]{Figs.~\ref{#1},\ref{#2},\ref{#3}}
\newcommand{\beq}{\begin{equation}}
\newcommand{\eeq}{\end{equation}}

\newcommand{\doublet}[3]{\:\left(\begin{array}{c} #1 \\#2
            \end{array} \right)_{#3}}
\newcommand{\vect}[2]{\:\left(\begin{array}{c} #1 \\#2
            \end{array} \right)}
\newcommand{\vectt}[3]{\:\left(\begin{array}{c} #1 \\#2 \\ #3
            \end{array} \right)}
\newcommand{\vectf}[4]{\:\left(\begin{array}{c} #1 \\#2 \\#3\\#4
            \end{array} \right)}
\newcommand{\matr}[4]{\left(\begin{array}{cc}
                   #1 &#2 \\
                   #3 &#4 \end{array} \right)}
\newcommand{\fracsm}[2]{{\textstyle\frac{#1}{#2}}}
\newcommand{\atopla}[2]{\displaystyle{{#1 \atop #2}}}

\newcommand{\D}{{\cal D}}
\newcommand{\K}{{\cal K}}
\newcommand{\NC}{N_{\rm CS}}
\newcommand{\Ncs}{N_{\rm CS}}
\newcommand{\SU}{$SU(2)~$}
\newcommand{\Pmax}{P_{max}}
\newcommand{\tr}{\,{\rm tr}\,}
\newcommand{\Tr}{\,{\rm Tr}\,}
\newcommand{\ldef}{=}
\newcommand{\rdef}{=}
\newcommand{\simlt}{\stackrel{<}{{}_\sim}}
\newcommand{\simgt}{\stackrel{>}{{}_\sim}}

\newcommand{\nuH}{\nu_{\SSS H}}
\newcommand{\nuF}{\nu_{\SSS F}}
\newcommand{\nuf}{\nu_{\SSS f}}
\newcommand{\nut}{\nu_{\SSS t}}
\newcommand{\nuHold}{\nu_{{\SSS H}\,{\rm old}}}

\newcommand{\op}[1]{{\bf \hat{#1}}}
\newcommand{\opr}[1]{{\rm \hat{#1}}}
\newcommand{\bra}[1]{\langle#1\vert}
\newcommand{\ket}[1]{\vert#1\rangle}
\newcommand{\lsim}{\mathrel{\lower 2pt\hbox{$\stackrel{<}{\SS\sim}$}}}
\newcommand{\gsim}{\mathrel{\lower 2pt\hbox{$\stackrel{>}{\SS\sim}$}}}

\newcommand{\mucr}{\mu_{\rm \SSS crit}}
\newcommand{\Vpot}{V_{\rm pot}}
\newcommand{\Vpotm}{V_{\rm pot}^\mu}
\newcommand{\Tkin}{T_{\rm kin}}
\newcommand{\rj}{r_{j+\frac{1}{2}}}
\newcommand{\omi}{\omega_{i+\frac{1}{2}}}
\newcommand{\phij}{\varphi_{j+\frac{1}{2}}}
\newcommand{\umin}{u_{\rm min}}
\newcommand{\umax}{u_{\rm max}}
\newcommand{\xmin}{x_{\rm min}}
\newcommand{\xmax}{x_{\rm max}}
\def\O{{\cal O}}

\def\theequation{\arabic{section}.\arabic{equation}}

 
\thispagestyle{empty}
\vspace*{-2.5cm}
\rightline{RUB-TPII-05/96}
\rightline{\today}
\vspace{1.5cm}
\begin{center} {\Large\bf
Spontaneous annihilation of high-density \\
matter in the electroweak theory}

\vspace{2.5cm}
{\large\bf
J\"org Schaldach$^\diamond$, Peter Sieber$^\diamond$, \\
Dmitri Diakonov$^*$\footnote{
\noindent diakonov@lnpi.spb.su},
and Klaus Goeke$^\diamond$\footnote{
\noindent goeke@hadron.tp2.ruhr-uni-bochum.de}} \\
\vspace{20 pt}
\noindent
{\small\it
$^\diamond$Inst. f\"ur Theor. Physik I\hskip -1.5pt I, Ruhr-Universit\"at
Bochum, D-44780 Bochum, Germany\\
$^*$St.~Petersburg Nuclear Physics Institute, Gatchina,
St.Petersburg 188350, Russia}
\end{center}
\vspace{1cm}
\abstract{In the presence of fermionic matter the topologically
distinct vacua of the standard model are metastable and can decay by
tunneling through the sphaleron barrier.  This process annihilates one
fermion per doublet due to the anomalous non-conservation of baryon and
lepton currents and is accompanied by a production of gauge and Higgs
bosons.  We present a numerical method to obtain local bounce solutions
which minimize the Euclidean action in the space of all configurations
connecting two adjacent topological sectors. These solutions determine
the decay rate and the configuration of the fields after the tunneling.
We also follow the real time evolution of this configuration and
analyze the spectrum of the created bosons. If the matter density
exceeds some critical value, the exponentially suppressed tunneling
triggers off an avalanche producing an enormous amount of bosons.}
\newpage

\section{Introduction}
\setcounter{equation}{0}

Baryon and lepton number violating processes in
the electroweak theory have been the
subject of many recent investigations.
They are due to the anomaly of the baryon and
lepton currents, discovered by 't Hooft \cite{tHooft},
and the non-trivial topological structure
of the electroweak theory. Faddeev \cite{Faddeev} and
Jackiw and Rebbi \cite{Jackiw} found that the potential energy
is periodic in a certain functional of the field,
the Chern--Simons number $N_{\rm CS}$, so that instead of one unique
vacuum there exist infinitely many field configurations with
zero potential energy, classified by integer values of $N_{\rm CS}$.

Each transition between vacua with $\Delta N_{\rm CS}=1$ is accompanied
by a change of the baryon and lepton number by one unit per fermion
generation. The vacua are
separated by an energy barrier, called sphaleron barrier
\cite{Dashen,Klinkhamer}, whose height is of the order of 10 TeV.
Under ordinary conditions the barrier can only be overcome by tunneling,
but the tunneling probability is suppressed by the factor
$\exp(-2 S_{\rm inst})\approx 10^{-153}$ with the instanton action
$S_{\rm inst}=8\pi^2/g^2\,,\ g\approx 0.67$, which means that the
process practically never happens.

Under special conditions, however, the fermion number violation rate might
well be significant. For example, a large temperature (of the order of $m_W$)
allows the system to cross the barrier classically \cite{Kuzmin,Arnold,rub3};
this process might have played a key role for the generation and conservation
of the baryon asymmetry in the early universe. The energy which is necessary
to overcome the barrier can possibly also be provided
by the incoming particles in a
collision if the particle energy is of the order of $10\sim 100$ TeV
\cite{Ringwald,Espinosa,McLerran,Mattis}.
Hence, fermion number violation
might be observable at future supercolliders.

In this paper we will investigate a third possibility to obtain
fermion number violation at a reasonable rate, namely in a surrounding of
high
density matter \cite{Rubakov,DiPet}. The mechanism of how the suppression
of the transition rate
is reduced is as follows: Matter of high density is described by a chemical
potential $\mu$, which is, at temperature zero, the energy up to which the 
Fermi-levels
are filled. A transition with $\Delta N_{\rm CS}=1$ creates a fermion which
has to be placed into the first free level, i.e.~it has energy $\mu$.
This energy must be added to the potential energy of the gauge and Higgs
bosons \cite{Rubakov,DiPet}:
\beq
\Vpotm=\Vpot+\mu\Ncs\ .
\la{vpotm}
\eeq
The extra term causes the previously degenerate vacua to become metastable
and the height of the barrier to be reduced
(see Fig.~1 where $\Vpotm$ is
plotted for a negative $\mu$) so that the tunneling probability increases.
At a certain critical value $\mu_{\rm crit}$,
the barrier disappears completely and the states at integer $N_{\rm CS}$
become unstable.
\begin{figure} [ht]
\frenchspacing
\centerline{
\epsfxsize=6.in
\epsfbox[85 455 553 700]{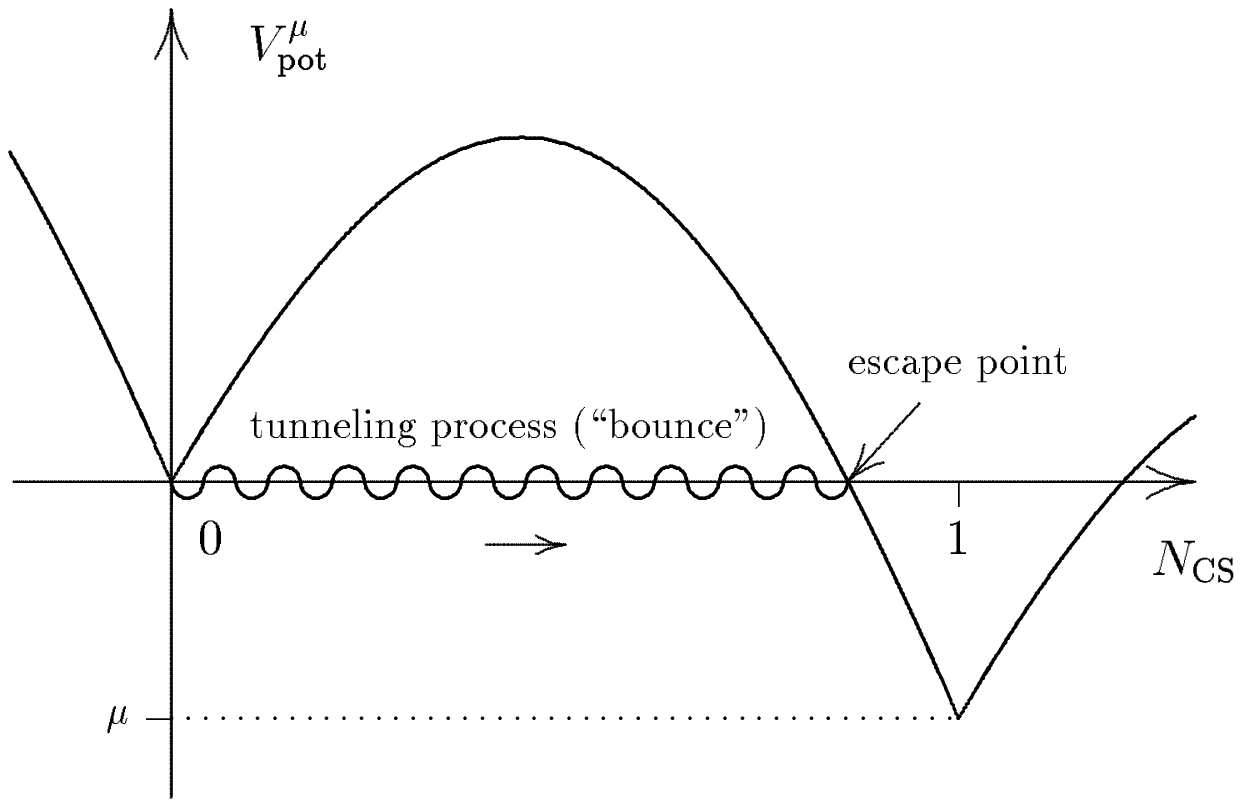}}
\nonfrenchspacing
\begin{quote}\begin{quote}
{\it Fig.~1:\/} Schematic plot of the tunneling process between
two topological sectors.
\end{quote}\end{quote}
\end{figure}

The decay rate of a metastable state per volume can be calculated by the
semi-classical WKB method. Following Coleman \cite{Coleman},
it is expressed in the form $\Gamma/V=B\exp(-A)$.
In order to find the exponent $A$ one has to solve the classical
Euclidean equations of motion, i.e.~one has to find the classical
motion of the system in the potential $-\Vpotm$ from some metastable
ground state (e.g.~a state with $\Ncs=0$, as indicated in Fig.~1) to a
configuration with the same potential energy which is on the other side
of the barrier.  We call this configuration ``escape point''.  The
exponent $A$ is twice the action of this motion, minimized over all
possible escape points.  In Euclidean space, after reaching the escape
point the system would move back the same way to the original ground
state, so that this process is called ``the bounce'' \cite{Coleman}.
Hence, the bounce represents the minimum of the Euclidean action in the
space of all possible paths from a metastable ground state to the other
side of the barrier.  The prefactor $B$ basically contains zero mode
factors and the determinant of small fluctuations about the bounce, but
in this work we will not be concerned with it and only compute the
exponent $A$.
 
It turns out that the chemical potential (and
hence the matter density) which is needed to obtain a reasonable decay
rate is quite large, so that it can hardly occur under normal
circumstances. Nevertheless, the results may still be relevant because
the decay rate is closely related to the rate of baryon and lepton
number violation at high particle energies \cite{DiPet}. Also, to our
knowledge, this is the first investigation of the metastable vacuum
decay in a real theory, beyond the so-called thin wall approximation;
and since it is a hard technical problem, our numerical
procedure might be useful in other applications, ranging from
spontaneous decay of heavy nuclei to inflationary scenarios of the
early universe.

In principle there are two different possibilities to find the bounce
numerically: Either one solves the equations of motion by
some initial value method like Runge--Kutta, or one considers the action
as a functional of the fields and minimizes it in the space of all
fields configurations. The first method seems to be unfeasible
because the potential $-\Vpotm$ has no lower bound so that a slight
deviation from the correct path will cause the system to fall
into some abyss of the potential $-\Vpotm$. We have therefore decided to take
the second
route. For the numerical implementation of the minimization
we use a discretization in space and time, based on a procedure presented
by Adler and Piran \cite{Adler}, so that the action becomes
a function of the values of the boson fields at the grid nodes.
At each step one considers the action as a function of
a certain field at a given point
and keeps the other values fixed. A single step of
a Newtonian algorithm is performed, then one takes another field
or moves to the next point
until one has performed a ``sweep'' through the whole lattice
and starts with the next one.

Related problems have already been treated in the literature, for
example in \cite{Rubakov1} the decay rate of a metastable state
was also calculated by minimizing the Euclidean action on a grid in
the context of technibaryons in the Skyrme model. In \cite{Koba1}
the Euclidean action of the electroweak theory is minimized, but
with respect to only a few parameters in the space of parametrized
functions. A method how to minimize the action of a model with scalar
fields was presented in \cite{Kusenko}.
A short letter about the present method has been published recently
\cite{rub4}.

Another aim of this work is to investigate the fate of the system
after the tunneling process has happened. It is known \cite{Coleman}
that the motion in Euclidean space does not only provide the
probability of the barrier penetration, but also yields the
most probable field configuration in which we will find
the system after the tunneling. This is just the escape point of
the bounce trajectory, i.e.~the configuration at the
other side of the barrier which belongs to the path in Euclidean space
with the least action. The potential energy of the system after the
tunneling is greater than the energy of the metastable minimum
in its current topological sector
(see Fig.~1). The difference $|\mu|$ is the energy of the annihilated
fermion which is now at the disposal of the boson fields.

The field configuration then performs a real time motion in Minkowski
space.
Knowing the escape point field configuration and the field velocities
(which are zero), one has to solve the standard Cauchi problem, which
we perform by straightforward integration of the second-order
differential equations.
Usually the system will fall towards the minimum in the sector of
the escape point, and eventually settles
at that minimum. The original energy $|\mu|$ of fermions from the Fermi surface
is converted into classical radiation of bosons. In 3+1
dimensions the amplitude of the outgoing wave package falls off as
$1/t$ so that the generally nonlinear
equations of motion can be linearized at large times $t$, and the outgoing
fields take the form of spherical waves.
In order to evaluate the particle content of
the multi-boson final state one has to carry out a Fourier
transformation of the fields after
their amplitudes got small enough. At this point we
basically follow the work of Hellmund and Kripfganz
\cite{Hellmund} (see also \cite{Cott}), who took a slightly disturbed
sphaleron as starting configuration, let it evolve in real time and
analyzed the resulting multi-boson state. Our main modifications are the
replacement of the sphaleron by the escape point of the bounce and the 
introduction of the chemical potential according to \eq{vpotm}.

In certain cases, however, when the fermion density is large enough
though less than the critical density $\mucr$, the real time
evolution of the fields is utterly different. The fields will not
quiet down at the minimum but ``splash'' over the {\em next} barrier
leading to a still lower minimum (at $\Ncs=2$ in Fig.~1), and so forth. As a
result the fermion sea will be completely ``dried out'', and a huge
amount of energy will be released in form of boson radiation.
Naively, one would think that such an avalanche happens only when the
fermion density exceeds the critical $\mucr$ when the system
is allowed to roll down classically. It is amusing that actually the
avalanche-like fermion annihilation can be triggered off by a spontaneous
tunneling process at the first stage.

The paper is organized as follows: In Section 2 we set up the model and
describe how matter of high density allows baryon and lepton number violating
processes. In Section 3 we present in detail our numerical procedure to
find the bounce trajectory. The classical motion after the tunneling is
investigated in Section 4. In Section 5 we give the numerical results of
our calculations, and finally we summarize our work in Section 6.

\section{Decay of high--density matter by barrier penetration}
\setcounter{equation}{0}

We consider the minimal version of the standard electroweak theory
with one Higgs doublet in the limit of vanishing Weinberg angle.
We work with dimensionless rescaled quantities, the corresponding
physical quantities can in general be obtained by multiplication with
appropriate powers of the gauge boson mass $m_W$. Sometimes this
factor is already included in the definition, for details see \cite{rub3}.
The bosonic part of the Lagrangian is
\beq
{\cal L}=\frac{m_W^4}{g^2}\biggl(
-\fracsm{1}{4}F_{\mu\nu}^a F^{a\,\mu\nu}
+\fracsm{1}{2}(D_\mu \Phi)^\dagger(D^\mu \Phi)
- \fracsm{1}{32}\nu^2(\Phi^\dagger\Phi-4)^2\biggr)
\la{Lagr}
\eeq
with the covariant derivative
$D_\mu=\partial_\mu -i A_\mu\;,\ A_\mu=\frac{1}{2}A_\mu^a\tau^a$,
the field strength $F_{\mu\nu}=\frac{1}{2}F_{\mu\nu}^a\tau^a=
i[D_\mu,D_\nu]\;,\ F_{\mu\nu}^a=\partial_\mu A_\nu^a-\partial_\nu A_\mu^a
+\e^{abc}A_\mu^b A_\nu^c$, and the Higgs doublet
$\Phi={\cphi^+ \choose \cphi^0}$. $\nu=m_H/m_W$ is the ratio of Higgs
and gauge boson masses; the $\tau^a$ are the Pauli matrices.
We work entirely in temporal gauge, $A_0=0$, which restricts possible
gauge transformations to time-independent ones,
\beq
A_i\to U(A_i+i\partial_i)U^\dagger\;,\quad\Phi\to U\Phi\qquad{\rm with}
\qquad U=U({\bf r})\in SU(2)\;.
\la{gaugetrans}
\eeq
The potential and kinetic energy and the Chern--Simons index are
\begin{eqnarray}
\Vpot&=&\frac{m_W}{g^2}\int
d^3{\bf r}\,\left[\fracsm{1}{4}(F^a_{ij})^2+\fracsm{1}{2}(D_i\Phi)^\dagger
(D_i\Phi)+\fracsm{1}{32}\nu^2(\Phi^\dagger\Phi-4)^2\right]\;, \nonumber\\
\Tkin&=&\frac{m_W}{g^2}\int
d^3{\bf r}\,\left[\fracsm{1}{2} (\dot A_i^a)^2+\fracsm{1}{2}\dot\Phi^\dagger
\dot\Phi\right]\;, \la{Eclass} \\
\Ncs&=&\frac{1}{16\pi^2}\int d^3{\bf r}\,\left[
\epsilon_{ijk}\left(A_i^a \partial_j A_k^a
+ \fracsm{1}{3} \epsilon_{abc} A_i^a A_j^b A_k^c\right)\right]\ . \nonumber
\end{eqnarray}
$\Ncs$ is only well-defined if the configuration space can be
identified with the sphere $S_3$, which requires the fields to be
continuous at infinity. We will always fix them to the trivial vacuum there
($A_i=0,\ \Phi={0 \choose 2}$). For vacuum configurations
$A_i=iU\partial_iU^\dagger$ with $U\in SU(2)$ (pure gauge), $\Ncs$ is
the integer winding number of the mapping $S_3\to SU(2)\sim S_3$.
Adjacent topological sectors are separated by energy barriers. The field
configurations which minimize the energy $\Vpot$ for given $\Ncs$ have been
calculated numerically by Akiba, Kikuchi, and Yanagida (AKY) \cite{AKY}.
\par
The fermions are coupled to the gauge and Higgs fields via the covariant
derivative and the Yukawa coupling, respectively. We do not consider them
explicitly, but note that due to the anomaly of the fermionic currents
their number is not
conserved, it varies with the Chern--Simons number of the classical
bosonic fields. For each doublet $(i)$ the fermion number changes as
\beq
\Delta N_i=\Delta\Ncs\;.
\la{deltaN}
\eeq
We assume to have a macroscopic amount of fermions of a very high
density in thermal equilibrium at zero (or very low) temperature.
There are $N_g(N_c+1)=12$ doublets, three leptonic and nine quark ones.
We describe them by the chemical potentials $\mu_i$ of the doublets;
since we have zero temperature, $\mu_i$ is the energy up to which the fermionic
levels are occupied (Fermi energy).
In a process connecting two adjacent vacua one level crosses the gap,
all others are shifted such that after the transition we have the same
spectrum again. But now one more level is occupied or depleted, depending
on whether the levels went up or down, hence the created or annihilated fermion
has the energy $\mu_i$. \Eq{deltaN} allows to include the change of the
fermionic energy into the bosonic energy functional. The new potential
energy is \cite{Rubakov,DiPet}
\beq
\Vpotm=\Vpot+\mu\Ncs
\la{vpotmu}
\eeq
with $\Vpot$ and $\Ncs$ from \ur{Eclass}.
$\mu=\sum_i\mu_i$ is the sum of the chemical potentials of the single
doublets. Obviously, we have fixed the zero-point of the energy to the
trivial vacuum with $\Ncs=0\,$. Besides, we can neglect the change
of the Fermi energy due to the creation or annihilation of
a small number of fermions.
\par
The additional term modifies the curvature of the potential around the
ground states of the topological sectors. A calculation of
the second derivative of $\Vpotm$ with respect to the fields
yields the following modes in momentum space \cite{Rubakov,DiPet}:
\beq
1+{\bf p}^2\;,\qquad\nu^2+{\bf p}^2\;,\qquad{\rm and}\qquad
1+{\bf p}^2\pm\frac{g^2\mu}{8\pi^2m_W}\,|{\bf p}|\;;
\la{vacmodes}
\eeq
besides, there is a zero mode due to gauge freedom. We see that there are
negative modes if $|\mu|$ exceeds the critical value
\beq
\mucr=\frac{16\pi^2}{g^2}\,m_W\;.
\la{mucrit}
\eeq
Therefore, at $|\mu|>\mucr$ the system may roll down
classically without any tunneling. A numerical calculation
\cite{rub} shows that
indeed the minimal energy barriers between the topological sectors
vanish in this case.
\par
We are mainly interested in the case $0<|\mu|<\mucr$, where the
topologically distinct minima with integer $\Ncs$ have
different energies $\Vpotm=\mu\Ncs$ and are separated by energy
barriers. Therefore, they are only local minima and are
thus metastable and can decay spontaneously
by quantum tunneling to the adjacent topological sector with lower
ground state energy. The energy difference $|\mu|$ between the two ground
states is the energy of the bosonic fields after the tunneling that
will eventually be carried away by the outgoing boson radiation.
\par 
As mentioned in the introduction, the tunneling rate per volume is of the
form \cite{Coleman}
\beq \Gamma/V=B\exp(-A) \la{rate} \eeq
where $A$ is
twice the Euclidean action of the bounce trajectory, which is the
classical path
that connects the decaying state and the turning point at the other
side of the inverted barrier and minimizes this action.
The prefactor
$B$ can be found from the small oscillation determinant about the
bounce, with one negative mode and the zero modes removed. $B$ also
contains the Jacobian
factors of the transformation groups which leave
the action invariant, i.e.~which correspond to the zero modes. The
factors coming from translational invariance in space and time are not
included in $B$ but in the left hand side of \ur{rate}, they lead to
the transition {\it rate per volume\/}.
\par In this paper we
concentrate on the calculation of $A$, we want to find the bounce
solution numerically. This means to find classical fields $A^a_i(t,{\bf
r})\,,\Phi(t,{\bf r})$ which represent a stationary point of the
Euclidean action 
\beq 
S_E=\int_{-\infty}^{t_0}dt\,m_W^{-1}(\Tkin+\Vpotm)\;.
\la{SE}
\eeq
At time $t=-\infty$ the fields form a metastable ground state with
integer $\Ncs$ and $\Tkin=0$; at $t=t_0$ the system reaches the
turning point at the other side of the potential valley (which is ---
considering the tunneling process --- the escape point at the other
side of the barrier). Here again $\Tkin=0$, and $\Ncs$ takes a non-integer
value belonging to the next topological sector. The bounce lasts
infinitely long because it starts from a ground state with
$\delta\Vpotm/\delta A^a_i=\delta\Vpotm/\delta \Phi=0\,$. Consequently,
$t_0$ is arbitrary, this corresponds to the translational invariance
in time mentioned above.
\par
As discussed previously, the choice of the temporal gauge still leaves the
freedom of time independent transformations.
Now we fix the gauge completely by demanding that at $t=-\infty$ the
fields start from the trivial vacuum $A^a_i=0\,,\ \Phi={0 \choose 2}$
with $\Ncs=0$ (and hence $\Vpotm=\Tkin=0$).
Besides, we choose $\mu<0$
so that the bounce moves from $\Ncs=0$ to the topological sector
with $\Ncs=1$.
\par
So far, we deal with 13 real functions depending on time $t$ and space
{\bf r}, nine from the gauge field $A^a_i$ and four from the complex
Higgs doublet $\Phi$. But we expect that our bounce solution possesses
higher symmetries than completely arbitrary fields.
Therefore we restrict our ansatz to fields having the spherical
symmetry of the sphaleron and the AKY configurations, which describe
the minimal energy barrier:
\begin{eqnarray}
A^a_i(t,{\bf r})&=&\epsilon_{aij}
n_j\,\frac{1-A(t,r)}{r}+(\delta_{ai}-n_an_i)\,
\frac{B(t,r)}{r}+n_an_i\,\frac{C(t,r)}{r}\;,\nonumber\\
\Phi(t,{\bf r})&=&\Bigl[H(t,r)+i G(t,r)\,{\bf n}\cdot{\bm\tau}\Bigr]
{{0 \choose 2}}
\la{hedge}
\end{eqnarray}
with $r=|{\bf r}|\;,\ {\bf n}={\bf r}/r\,$.
This reduces the effort to five real functions depending on $t$ and $r$.
The free choice of the origin corresponds to the spatial translational
invariance discussed after \eq{rate}.
Gauge transformations within this ansatz are given by
\beq
U(r)=\exp\Bigl[{\bf n}\cdot{\bm\tau}\,iP(r)\Bigr]\;,
\la{hedgegauge}
\eeq
they transform the fields as
\begin{eqnarray}
A(t,r)&\to&A(t,r)\cos 2P(r)-B(t,r)\sin 2P(r)\,, \nonumber \\
B(t,r)&\to&B(t,r)\cos 2P(r)+A(t,r)\sin 2P(r)\,, \nonumber \\
C(t,r)&\to&C(t,r)+2rP'(r)\,,                      \la{hedgau} \\
H(t,r)&\to&H(t,r)\cos P(r)-G(t,r)\sin P(r)\,, \nonumber \\
G(t,r)&\to&G(t,r)\cos P(r)+H(t,r)\sin P(r)\,. \nonumber
\end{eqnarray}
We will use this transformation later to adjust the numerical solutions to our 
choice of the gauge.
\par
As mentioned above, we choose a gauge which yields the trivial vacuum
at $t=-\infty$. In this gauge the fields $A^a_i,\;\Phi$ of the bounce solution
are continuous and differentiable everywhere and have
finite potential and kinetic energy, because this is true for our starting 
point and will not be changed during the evolution governed
by the Euclidean equations of motion. This requires the following
behavior of the radial functions at $r=0\,$:
\begin{eqnarray}
A(t,r)&=&1+a_2(t)\,r^2+\O(r^3)\,, \nonumber\\
B(t,r)&=&b_1(t)\,r+\O(r^3)\,, \nonumber\\
C(t,r)&=&b_1(t)\,r+\O(r^3)\,, \la{origin}\\
H(t,r)&=&h_0(t)+h_2(t)\,r^2+\O(r^3)\,, \nonumber\\
G(t,r)&=&g_1(t)\,r+g_2(t)\,r^2+\O(r^3)\,. \nonumber
\end{eqnarray}
\par
The numerical determination of the bounce is performed by finding
a stationary point of the Euclidean action directly, without using the
equations of motion.
In our spherical ansatz \ur{hedge}, $S_E$ is a functional of the 
five functions
$A,B,C,H,G$ of \eq{hedge}, depending on radial distance $r$ and time $t$.
The bounce has infinite extension as well in space as in time,
but we can introduce new variables $x$ and $u$ which cover
only finite intervals, for example
\beq
r(x)=\lambda_r\arctan\left(\fracsm{\pi}{2}\,x\right)
\qquad{\rm and}\qquad
t(u)=\lambda_t\arctan\left(\fracsm{\pi}{2}\,u\right)\ ,
\la{subst}
\eeq
and new profile functions depending on $x$ and $u$.
Using ansatz \ur{hedge} and the substitution \ur{subst} we get
\begin{eqnarray}
S_E&=&\frac{S_{\rm inst}}{2\pi}\int_{-1}^{u_0}du\int_0^1 dx
\,\frac{1}{\om\vp}\biggl[\om^2\Bigl(\dot A^2+\dot B^2+\fracsm{\dot C^2}{2}
+2r^2(\dot H^2+\dot G^2)\Bigr)
\nonumber \\
&&{}+\Bigl(\vp A'+\fracsm{BC}{r}\Bigr)^2+\Bigl(\vp B'-\fracsm{AC}{r}\Bigr)^2
+2r^2\Bigl(\vp H'+\fracsm{GC}{2r}\Bigr)^2
\nonumber \\
&&{}+2r^2\Bigl(\vp G'-\fracsm{HC}{2r}\Bigr)^2
+\fracsm{1}{2r^2}(A^2+B^2-1)^2
+H^2\Bigl((A-1)^2+B^2\Bigr)
\nonumber \\
&&{}+G^2\Bigl((A+1)^2+B^2\Bigr)-4BGH+\fracsm{1}{2}\nu^2 r^2(H^2+G^2-1)^2
\nonumber \\
&&{}+2\rho\Bigl(\fracsm{C}{r}(A^2+B^2-1)+\vp BA'-\vp(A-1)B'\Bigr)\biggr]
\la{SEhedge}
\end{eqnarray}
with
\beq
\rho=\frac{\mu}{\mucr}\;,\qquad\vp=\vp(x)=\left(\frac{dr}{dx}\right)^{-1}\;,
\qquad\om=\om(u)=\left(\frac{dt}{du}\right)^{-1}\;,
\eeq
and the dot and prime mean $\fracsm{d}{du}$ and $\fracsm{d}{dx}$, 
respectively. In the next section we describe how we find stationary 
points of the functional \ur{SEhedge} numerically.

\section{Numerical determination of the bounce trajectory}
\setcounter{equation}{0}

Our way to find a stationary point of $S_E$ is the use of a relaxation
method which was discussed by Adler and Piran in great detail \cite{Adler}.
The functional \ur{SEhedge} is put on a two-dimensional grid of size
$(n_u+1)\times(n_x+1)$. We distinguish the full node grid with points
\beq
(u_i,\,x_j)=(\umin+i\,\Delta u,\,\xmin+j\,\Delta x)\;,\qquad
\left({i=0,\dots,n_u\atop j=0,\dots,n_x}\right)
\la{fulln}
\eeq
and the half node grid with points
\beq
(u_{i+\frac{1}{2}},\,x_{j+\frac{1}{2}})=(\umin+(i+\fracsm{1}{2})\,\Delta u,\,
\xmin+(j+\fracsm{1}{2})\,\Delta x)\;,\quad
\left({i=0,\dots,n_u-1\atop j=0,\dots,n_x-1}\right)
\la{halfn}
\eeq
with
\beq
\Delta u=\frac{\umax-\umin}{n_u}
\qquad\mbox{and}\qquad
\Delta x=\frac{\xmax-\xmin}{n_x}\;.
\eeq
For the substitution \ur{subst} the values are
e.g.~$\umin=-1\,,\ \umax=0$ (for $t_0=0$), $\xmin=0\,,\ \xmax=1\,$.
The five profile functions are put on the full node grid, the
radial distance $r$ and the derivatives $\omega$ and $\vp$ on the
half node grid. We introduce the notation
\beq
A^i_j=A(u_i,x_j)\;,\qquad
\rj=r(x_{j+\frac{1}{2}})
\la{nodes}
\eeq
and accordingly for the other functions.
The use of the half node grid prevents the calculation of expressions
like $r$ or $1/r$ at the boundaries $r=\infty,0$.
\par
There are many different ways to put an integral as \ur{SEhedge} on a
grid, so that in the limit $n_u,n_x\to\infty$ the original functional
is restored again. Therefore it is rather important to choose a
discretization that is suitable for numerical treatment. Basically, we
followed the suggestions of \cite{Adler} here, but to fix the details
we had to try and compare different ansatzes. For example, the property
of some terms to vanish at the origin and cancel the $1/r$
divergence must not be lost by the discretization. It showed that
terms as
$(\vp A'+\frac{BC}{r})$ should be discretized before being squared,
and that instead of $C(t,r)$ the function
\beq
D(u,x)\equiv \frac{C(u,x)}{r(x)}
\la{DCr}
\eeq
should be used, i.e.~put on the full node grid as $D^i_j$.
We use the following discretization for $S_E$, \eq{SEhedge},
written here as sum over contributions from grid cells centered
at half node points
$(u_{i+\frac{1}{2}},\,x_{j+\frac{1}{2}})\,$:
\newlength{\arraycolsepnormal}
\setlength{\arraycolsepnormal}{\arraycolsep}
\setlength{\arraycolsep}{2pt}
\begin{eqnarray}
S_E^{\rm grid}&=&\frac{S_{\rm inst}}{2\pi}\,
\sum_{i=0}^{n_u-1}\sum_{j=0}^{n_x-1}\,\frac{\Delta u\,\Delta x}{\omi\phij}\;
\Biggl[\frac{\omi^2}{2\Delta u^2}\,\biggl((A^{i+1}_j-A^i_j)^2
\la{SEgrid} \\
&&\quad{}+(A^{i+1}_{j+1}-A^i_{j+1})^2
+(B^{i+1}_j-B^i_j)^2+(B^{i+1}_{j+1}-B^i_{j+1})^2
\nonumber \\
&&\quad{}+2\rj^2\Bigl((H^{i+1}_j-H^i_j)^2+(H^{i+1}_{j+1}-H^i_{j+1})^2
+(G^{i+1}_j-G^i_j)^2
\nonumber \\
&&\quad{}+(G^{i+1}_{j+1}-G^i_{j+1})^2
+\fracsm{1}{4}(D^{i+1}_j-D^i_j)^2
+\fracsm{1}{4}(D^{i+1}_{j+1}-D^i_{j+1})^2\Bigr)\biggr)
\nonumber \\
&+&\fracsm{1}{2}\,\Biggl\{
\Bigl(\frac{\phij}{\Delta x}(A^i_{j+1}-A^i_j)
+\fracsm{1}{2}(B^i_jD^i_j+B^i_{j+1}D^i_{j+1})\Bigr)^2
\nonumber \\
&&{}+\Bigl(\frac{\phij}{\Delta x}(B^i_{j+1}-B^i_j)
-\fracsm{1}{2}(A^i_jD^i_j+A^i_{j+1}D^i_{j+1})\Bigr)^2
\nonumber \\
&&{}+2\rj^2\Bigl(\frac{\phij}{\Delta x}(H^i_{j+1}-H^i_j)
+\fracsm{1}{4}(G^i_jD^i_j+G^i_{j+1}D^i_{j+1})\Bigr)^2
\nonumber \\
&&{}+2\rj^2\Bigl(\frac{\phij}{\Delta x}(G^i_{j+1}-G^i_j)
-\fracsm{1}{4}(H^i_jD^i_j+H^i_{j+1}D^i_{j+1})\Bigr)^2
\nonumber \\
&&{}+\frac{1}{4\rj^2}\biggl(({A^i_j}^2+{B^i_j}^2-1)^2
+({A^i_{j+1}}^2+{B^i_{j+1}}^2-1)^2\biggr)
\nonumber \\
&&+\fracsm{1}{2}\biggl(
{H^i_j}^2\Bigl((A^i_j-1)^2+{B^i_j}^2\Bigr)
+{H^i_{j+1}}^2\Bigl((A^i_{j+1}-1)^2+{B^i_{j+1}}^2\Bigr)
\nonumber \\
&&\quad{}+{G^i_j}^2\Bigl((A^i_j+1)^2+{B^i_j}^2\Bigr)
+{G^i_{j+1}}^2\Bigl((A^i_{j+1}+1)^2+{B^i_{j+1}}^2\Bigr)
\nonumber\\
&&\quad{}-4B^i_jG^i_jH^i_j-4B^i_{j+1}G^i_{j+1}H^i_{j+1} \biggr)
\nonumber \\
&&{}+\fracsm{1}{4}\nu^2\rj^2\biggl(({H^i_j}^2+{G^i_j}^2-1)^2
+({H^i_{j+1}}^2+{G^i_{j+1}}^2-1)^2 \biggr)
\nonumber \\
&&{}+\rho\biggl(D^i_j({A^i_j}^2+{B^i_j}^2-1)
+D^i_{j+1}({A^i_{j+1}}^2+{B^i_{j+1}}^2-1) \biggr)
\nonumber \\
&&{}+\rho\,\frac{\phij}{\Delta x}\biggl((A^i_{j+1}-A^i_j)(B^i_j+B^i_{j+1})
-(A^i_j+A^i_{j+1}-2)(B^i_{j+1}-B^i_j) \biggr)\Biggr\}
\nonumber \\
&+&\fracsm{1}{2}\,\Biggl\{\quad i\to i+1\quad\Biggr\} \Biggr]\,.
\nonumber
\end{eqnarray}
\setlength{\arraycolsep}{\arraycolsepnormal}
\par
Typically, we used grids of the size $n_u=n_x=80$, in which case the
profile functions are represented by $5\cdot 81\cdot 81=32805$ points.
To find a stationary point of $S_E^{\rm grid}$ now means to find a
configuration which fulfills the equations
\beq
\frac{\partial S_E^{\rm grid}}{\partial A^i_j}=0\;,\quad
\frac{\partial S_E^{\rm grid}}{\partial B^i_j}=0\;,\quad
\frac{\partial S_E^{\rm grid}}{\partial D^i_j}=0\;,\quad
\frac{\partial S_E^{\rm grid}}{\partial H^i_j}=0\;,\quad
\frac{\partial S_E^{\rm grid}}{\partial G^i_j}=0
\la{statpoint}
\eeq
for all $i$ and $j$.
\par
The solution is found iteratively until the configuration satisfies
\eq{statpoint} sufficiently well. In each step, only one single number
$A^i_j\,,\;B^i_j\,,\;D^i_j\,,\;H^i_j\,$ or $G^i_j\,$ for certain $i,j$ is\
 changed,
all others are held constant.
``Sweeping'' over the grid, we modify the field parameters one after
another, where the order is only of minor importance. (We changed the
five functions for given $i,j$ and varied $j$ for fixed $i$.)
The change of e.g.~$A^i_j$ is governed only by the according equation
$\partial S_E^{\rm grid}/\partial A^i_j=0$, all other equations are not
taken into account. We perform the first step of a Newtonian algorithm
that would
converge against the solution of that equation, i.e.~we change
the field parameter as
\beq
{A}^i_j\quad\longrightarrow\quad A^i_j-\kappa\,
\frac{\partial S_E^{\rm grid}/\partial A^i_j}
{\partial^2 S_E^{\rm grid}/\partial{A^i_j}^2}\,,
\la{newton}
\eeq
where in general for the damping parameter we choose $\kappa=1$.
The necessary partial derivatives in \eq{newton} must be calculated
from \ur{SEgrid}, we do not write them explicitly here.
\par
Unfortunately, the bounce solution is not a local minimum, but only a
saddle point of the Euclidean action:
The problem is that because of
the term $\mu\Ncs$ the Euclidean potential $-\Vpotm$ is not bound from above,
it can take positive values. Due to energy conservation the bounce
itself between $\Ncs=0$ and 1 cannot have
positive potential energy $-\Vpotm$ at any time,
its total energy $E_{\rm tot}=\Tkin-\Vpotm$ is constant and zero (Fig.~1).
But in its vicinity,
one can construct paths which have positive potential $-\Vpotm$ for
some time and which give a {\it lower} action than the bounce.
We found that unrestricted sweeps according to
\ur{newton} always lead to configurations of that kind. Once there,
the system quickly evolves to enormously high winding numbers $\Ncs$
and an unlimitedly decreasing negative action. Thus, in order to avoid
this instability, we have to prevent the system from acquiring positive
potential $-\Vpotm$. We did so by choosing an initial configuration
with non-positive potential, and then rejecting all steps
\ur{newton} which would yield a positive $-\Vpotm$. If a step is not accepted,
it is tried again with $\kappa$ divided by 2; this is repeated up to
five times before the step is completely rejected for the current sweep.
Unfortunately, this method requires to calculate $\Vpotm$ (or actually the
change of $\Vpotm$) for the current time slice after each single step,
which slows down the algorithm considerably.
There are more sophisticated ways
to take into account invariances like energy conservation \cite{Kusenko},
but in our case the simple remedy proved to be the most effective.
Within the restriction $-\Vpotm\le0$ the bounce locally minimizes the action.
Numerically, we find indeed that the action decreases monotonically and
converges to some limit.
\par
As starting configuration we usually take an instanton-like configuration,
\begin{eqnarray}
A(t,r)&=&\cos\beta-2\,\frac{rt\sin\beta+r^2\cos\beta}{r^2+t^2+\lambda^2}\,,
\la{iniconf} \\
B(t,r)&=&-\sin\beta-2\,\frac{rt\cos\beta-r^2\sin\beta}{r^2+t^2+\lambda^2}\,,
\nonumber \\
D(t,r)&=&-\frac{\lambda^2}{r(r^2+\lambda^2)}\,
\Bigl(\beta+\frac{2rt}{r^2+t^2+\lambda^2}\Bigr)\,, \nonumber \\
H(t,r)&=&1-\fracsm{1}{2}\Bigl(1+\frac{t}{t^2+\lambda^2}\Bigr)
\Bigl(1+\cos\frac{\pi r}{\sqrt{r^2+\lambda^2}}\Bigr)\,,
\nonumber \\
G(t,r)&=&\fracsm{1}{2}\Bigl(1+\frac{t}{t^2+\lambda^2}\Bigr)
\sin\frac{\pi r}{\sqrt{r^2+\lambda^2}}\,,
\nonumber \\
\mbox{with}\qquad
\beta&=&\beta(t,r)=\frac{2r}{\sqrt{r^2+\lambda^2}}\,
\Bigl(\arctan\frac{t}{\sqrt{r^2+\lambda^2}}+\fracsm{\pi}{2}\Bigr)\,.
\nonumber
\end{eqnarray}
The size parameter $\lambda$ is chosen between 2 and 4.
This configuration has potential
$-\Vpotm=0$ at $t=-\infty$ and $-\Vpotm=-\mu>0$ at $t=\infty\,$; we choose
the boundary $\umax$ of the grid such that at $t(\umax)$ the potential
crosses the zero line. Hence for $\umin<u<\umax$ we have $-\Vpotm<0\,$.
\par
According to our gauge fixing and to \eq{origin} the fields are
fixed at some boundaries of our grid. For $u=\umin\ (t=-\infty)$ or
$x=\xmax\ (r=\infty)$
we have $A=H=1\,,\ B=D=G=0\,$, and for the origin $x=\xmin\ (r=0)$ we know
$A=1\,,\ B=G=0\,$. Accordingly, the following field parameters are held
fixed:
\begin{eqnarray}
&&A^0_j=A^i_{n_x}=A^i_0=1\,, \nonumber \\
&&B^0_j=B^i_{n_x}=B^i_0=0\,, \nonumber \\
&&G^0_j=G^i_{n_x}=G^i_0=0\,, \la{boundary} \\
&&D^0_j=D^i_{n_x}=0\,, \nonumber \\
&&H^0_j=H^i_{n_x}=1\,. \nonumber
\end{eqnarray}
\par
Unfortunately, if one only sweeps over the grid using \ur{newton}, the\
   outcome
will not be a reasonable bounce solution, but a rather discontinuous and\
   unusable
configuration. Therefore the computational procedure cannot be run from start
to end automatically, but requires the controlling and regulating of the user
once in a while, which makes the work a tedious and long lasting one.
The following problems arise:
\par
The Euclidean action \ur{SEhedge} contains no term $D'=\partial D/\partial
  x\,$.
Hence, adjacent field parameters $D^i_j$ and $D^i_{j+1}$
are only weakly coupled and the $D$ field
easily ceases to be smooth. Especially close to the origin the field is
rather unstable, therefore we do not use \ur{newton} to fix $D^i_0$ and
$D^i_1\,$, but adjust them to a linear extrapolation through
$D^i_2$ and $D^i_3\,$.
Nevertheless, in order to keep the fields reasonably smooth, we are
forced to smooth them by an averaging procedure from time to time, where each
field parameter is replaced by the average value of itself
and some of its neighbors,
with weight factors according to their distance. Certainly this disturbs the
minimization algorithm and generally leads to a higher action again. But after
some further sweeps the action is down to its earlier
value again, and the fields are smoother now.
\par
Another problem is the following: Close to $u=\umin\ (t=-\infty)$ the
factor $\omega=du/dt$ which governs the kinetic energy terms is rather small
so that adjacent time slices are only weakly coupled. Therefore,
the fact that
we fixed the fields to the trivial vacuum at the $u=\umin$ boundary
hardly influences the configuration at larger times. Instead, we usually see
that going from large $u$ towards $\umin$ the configuration continuously
approaches a {\it non-trivial} vacuum state, and then close to
$\umin$ the fields show a
discontinuous step from this non-trivial vacuum to the trivial one. We get
rid of this step by performing a gauge transformation of the kind \ur{hedgau}
which converts the non-trivial vacuum at the edge of the step to the
trivial one. This gauge transformation is applied to all time slices
except the ones close to $\umin$ where the fields are already trivial.
\par
Apart from the above manipulations which are necessary to keep the
configuration in an acceptable shape, we also have the
possibility to accelerate the convergence of the algorithm. The solution
has to obey the energy conservation law $E_{\rm tot}=\Tkin-\Vpotm=0$. Given
some arbitrary configuration for which this is not the case, one can find
a different time parameterization such that energy conservation
is fulfilled. Practically, we leave the values $t(u_i)$ of the
times at the grid nodes fixed and determine the fields of the reparametrized
configuration at those times $t(u_i)$ by interpolation.
By this operation one can gain a considerable decrease of the
action without performing minimization sweeps.
\par
Finally we remark that the grid size is not completely fixed, but it is
adapted to the status of the minimization. Usually we start with a size of 
$41\times 41$
nodes, and only when we are already close to the solution we double the
grid to $81\times 81$ points. Moreover, we observe that during the sweeps
the bounce, especially its escape point where the potential reaches zero again, 
slowly moves towards smaller times.
In principle, the time scale is arbitrary, as discussed above, but for
numerical reasons a given configuration has slightly lower action when it
is shifted to smaller times. We find that the energy at a few points next
to $\umax$ becomes zero, so that the escape
point does not longer coincide with $\umax$. If the number of these points
gets too large, we throw away the part of the configuration beyond the
escape point, which results in a lower $\umax$ and $n_u$. If $n_u$ becomes
too small, we double the grid in the time dimension only, i.e.~we double
$n_u$ but leave $n_x$. For the final configuration we adjust the time scale
such that the escape point is set on the origin, i.e.~$t_0=0$.

Practically, the bounce trajectory is found by switching between the
minimization sweeps and one of the manipulations described above.
If any and which of the regulations should be performed has to be decided
by looking at the actual field configuration. At first sight this
procedure seems to be subjective and non-reproducible, but let us
remark that before we stop the program, the last manipulation is always
followed by at least 500 sweeps.  Moreover we have always checked that
the final configuration fulfills the equations of motion with excellent
accuracy.

\section{The real time evolution after the tunneling}
\setcounter{equation}{0}

In this section we show how we investigate the behavior of the boson
field configuration after the barrier penetration and how we analyze
the particle content of the state. We follow basically the procedure
presented in \cite{Hellmund}, but instead of the sphaleron we take
the escape point of the bounce as starting configuration.

The potential energy of the escape point is larger by $|\mu|$ than
the potential energy of the ground state in the corresponding
topological sector.
Hence, the system performs a motion in the real time Minkowski space.
The equations of motion within the spherical ansatz \ur{hedge} are
\begin{eqnarray}
\ddot{A} &=& \Bigl(A'+\fracsm{BC}{r}\Bigr)'-\fracsm{A}{r^2}(A^2+B^2-1)
      +\Bigl(\fracsm{C}{r}+2\rho\Bigr)\,\Bigl(B'-\fracsm{AC}{r}\Bigr)
    \nonumber \\ &&\hspace{5.0cm} {}-A(H^2+G^2)+H^2-G^2\,, \nonumber \\
\ddot{B} &=& \Bigl(B'-\fracsm{AC}{r}\Bigr)'-\fracsm{B}{r^2}(A^2+B^2-1)
      -\Bigl(\fracsm{C}{r}+2\rho\Bigr)\,\Bigl(A'+\fracsm{BC}{r}\Bigr)
    \nonumber \\ &&\hspace{5.5cm} {}-B(H^2+G^2)+2HG\,, \nonumber \\
\ddot{C} &=& \fracsm{2A}{r}\Bigl(B'-\fracsm{AC}{r}\Bigr)
    -\fracsm{2B}{r}\Bigl(A'+\fracsm{BC}{r}\Bigr) -C(H^2+G^2) \nonumber \\
    &&\hspace{3.2cm} {}+2r(HG'-H'G)-\fracsm{2\rho}{r}(A^2+B^2-1)\,, \nonumber
   \\
\ddot{H} &=& \fracsm{1}{r}(rH)'' + \fracsm{1}{2r^2}(CG+C'Gr+2CG'r)
      +\fracsm{1}{r^2}(AH+BG) \nonumber \\
  &&\hspace{1.5cm} {} - \fracsm{H}{2r^2}\Bigl(1+A^2+B^2+\fracsm{C^2}{2}\Bigr)
  -\fracsm{\nu^2}{2}H(H^2+G^2-1)\,, \nonumber \\
\ddot{G} &=& \fracsm{1}{r}(rG)'' - \fracsm{1}{2r^2}(CH+C'Hr+2CH'r)
      +\fracsm{1}{r^2}(BH-AG) \nonumber \\
  &&\hspace{1.5cm} {} - \fracsm{G}{2r^2}\Bigl(1+A^2+B^2+\fracsm{C^2}{2}\Bigr)
  -\fracsm{\nu^2}{2}G(H^2+G^2-1)\,, \label{eqnmotion}
\end{eqnarray}
where the dot means the derivative with respect to the time $t$ and the prime
with respect to the radial coordinate $r$.

These equations and the condition that at $t=0$ the system
starts at the escape point of the bounce with kinetic energy zero
(i.e.~$\dot{A}(0)=\dot{B}(0)=\dot{C}(0)=\dot{H}(0)=\dot{G}(0)=0$)
form a Cauchi problem which is solved by
direct integration. To do this we discretize the Minkowskian action in the
same
way as the Euclidean action, which means we take \eq{SEgrid} and
reverse the sign of the terms stemming from the potential energy.
Moreover we identify $t(u)=u$ instead of \eq{subst}, so that $\omega(u)=1$,
but we still keep the relation between $r$ and $x$ of \ur{subst}.
We obtain discretized equations of motion by deriving
the discretized Minkowskian action with respect to the field coordinates
$A^i_j,\,\ldots,\,G^i_j$. These equations are solved for the variables
$A^{i+1}_j,\,\ldots,\,G^{i+1}_j$ so that we obtain the
fields with time index $i$+1 as a function of those with indices $i$
and $i$--1. By iterative application of these equations it is then possible
to evaluate the propagation of the system from the initial configuration
at $t=0$ to arbitrary positive times.

The grid of the discretization can be much
more dense here than in the case of the Euclidean problem. Usually we
take
800 steps per time unit and 3000 grid nodes in the interval from
$x=0$ to $x=1$. We checked that the results are stable with respect to
a further increase of these parameters. The total time how long
we follow the propagation of the fields is typically around 20 to 30
(in units of $m_W^{-1}$).

As in the Euclidean case, special care has to be taken in order to treat
the fields close to the origin $r=0$ adequately.
If the numerical solution does not exactly fulfill the expansion given
by \eq{origin}, some terms of the r.h.s.~of \eq{eqnmotion}
become singular. Hence, even a slight deviation from this expansion
increases rather quickly with $t$ and finally
results in a strong divergence of the fields
close to $r=0$. Since small numerical errors will always cause this to
happen we cannot take the discretized equations of motion close to the
origin, but we rather impose the behavior of \ur{origin} by hand. For
about
the first 50 of the 3000 points we determine the fields not by the iteration
method but by \eq{origin} where the coefficients $a_2(t)$ etc.~are chosen
such that the functions $A,\,\ldots,\,H$ are continuous and differentiable
at the matching point between the numerical solution and the fit \ur{origin}.

We find that for most sets of parameters $\nu$ and $\rho$ after some time
the fields perform small oscillations about some vacuum configuration
$\bar{A}(r),\,\ldots,\,\bar{G}(r)$ in the topological
sector of the escape point (which is the one with $\Ncs=1$ for our choice of
gauge). An indication for this behavior is that $\Ncs$,
$\Vpotm$, and $\Tkin$ do not change any more with $t$, the energies take
constant values $\pm\frac{|\mu|}{2}$ according to the virial theorem.

The Fourier analysis of the small oscillations is greatly
simplified if the vacuum about which the fields are fluctuating is the
trivial one. For this reason we perform a time independent gauge
transformation of the type \ur{hedgau} which transforms the configuration
$\bar{A}(r),\,\ldots,\,\bar{G}(r)$ into the trivial vacuum.
In order to determine the configuration
$\bar{A}(r),\,\ldots,\,\bar{G}(r)$, we start the propagation
of the fields with the original escape point and average the fields:
\beq
\bar{A}(r)=\frac{1}{t-t_s}\int^t_{t_s}dt'\,A(t',r)\,,
\label{average}\eeq
and equivalently for the other profiles. $t_s$ is some time where the system
already performs small oscillations. We checked that
$\bar{A}(r),\,\ldots,\,\bar{G}(r)$ are independent of $t$ if $t$ is large
enough and that they in fact represent a vacuum configuration with $\Ncs=1$.

The gauge transformation which transforms
$\bar{A}(r),\,\ldots,\,\bar{G}(r)$ into the trivial vacuum
changes the Chern--Simons number by $\Delta\Ncs=-1$. We
apply it to our starting configuration, the escape point, which hereby gets
a Chern--Simons number between $-1$ and 0 and the potential energy
$\Vpotm=|\mu|$. Now we start the propagation again with the
starting configuration in the new gauge. 
By performing the same averaging process again,
we can check that the system now indeed fluctuates about the trivial
vacuum with excellent accuracy, which proves that the gauge invariance
is correctly reproduced in our numerics.

For the following we assume that the system has reached the status where it
   can
be described by small fluctuations about the trivial vacuum. We denote
these fluctuations with small letters:
\begin{eqnarray}
&&A(t,r)=1+a(t,r)\,,\hspace{1cm} B(t,r)=b(t,r)\,,\hspace{1cm}
   C(t,r)=c(t,r)\,,
\nonumber \\
&&H(t,r)=1+h(t,r)\,,\hspace{1cm} G(t,r)=f(t,r)\,, \label{fluct}
\end{eqnarray}
and expand the energy to second order in the fluctuations:
\begin{eqnarray}
\Tkin^{(2)}&=&\frac{4\pi m_W}{g^2}\int^\infty_0 dr\,\Bigl(\dot{a}^2+\dot{b}^2
   +\fracsm{\dot{c}^2}{2} + 2r^2 (\dot{h}^2+\dot{f}^2)\Bigr)\,,
\nonumber \\
{\Vpotm}^{(2)}&=&\frac{4\pi m_W}{g^2}\int^\infty_0
   dr\,\biggl[\fracsm{2a^2}{r^2}
  +a'^2+\Bigl(b'-\fracsm{c}{r}\Bigr)^2+a^2+b^2+\fracsm{c^2}{2}+4f^2-4bf
   \nonumber \\
  &&\hspace{1.3cm} -2rcf'+2r^2(h'^2+f'^2)+2\nu^2r^2h^2
  +2\rho\Bigl(a'b-b'a+\fracsm{2ac}{r}\Bigr)\biggr]\,, \nonumber \\
E_{\rm tot}^{(2)}&=&\Tkin^{(2)}+{\Vpotm}^{(2)}\,.
\label{en2order}\end{eqnarray}
We found that ${\Vpotm}^{(2)}$ and $\Vpotm$ coincide up to a deviation of less 
than
$1\%$ for large $t$ which indicates that the system has perfectly linearized.
The potential in second order leads to the linear equations of motion
\begin{eqnarray}
\ddot{a} &=& a''-a\Bigl(1+\fracsm{2}{r^2}\Bigr)
       +2\rho\Bigl(b'-\fracsm{c}{r}\Bigr)\,, \nonumber \\
\ddot{b} &=& \Bigl(b'-\fracsm{c}{r}\Bigr)'-b + 2f-2\rho a'\,,
 \nonumber \\
\ddot{c} &=& \fracsm{2}{r}\Bigl(b'-\fracsm{c}{r}\Bigr) -c+2rf'
    -4\rho\fracsm{a}{r}\,, \nonumber \\
r\ddot{h} &=& (rh)'' -\nu^2 rh\,, \nonumber \\
r\ddot{f} &=& (rf)'' -\fracsm{1}{2r}(rc)'+\fracsm{b}{r} -\fracsm{2f}{r}\,.
\label{eqnsecond}\end{eqnarray}
In the case $\rho\ne 0$, for an arbitrary fixed momentum $k$ the
solution is
\begin{eqnarray}
a_k(t,r)&=&\frac{3}{2}\Bigl(\beta_1^k(t)-\beta_2^k(t)\Bigr)\,rj_1(kr)\,,
\nonumber \\
b_k(t,r)&=&\Bigl(\beta_0^k(t)+\beta_1^k(t)+\beta_2^k(t)
         + \zeta^k(t)\Bigr)\,rj_0(kr) \nonumber \\
 &&{}+\Bigl(\beta_0^k(t)-\fracsm{1}{2}\beta_1^k(t)
 -\fracsm{1}{2}\beta_2^k(t) + \zeta^k(t)\Bigr)\,rj_2(kr)\,,\nonumber \\
c_k(t,r)&=&\Bigl(\beta_0^k(t)+\beta_1^k(t)+\beta_2^k(t)
         + \zeta^k(t)\Bigr)\,rj_0(kr) \nonumber \\
 &&{}-2\,\Bigl(\beta_0^k(t)-\fracsm{1}{2}\beta_1^k(t)-\fracsm{1}{2}
\beta_2^k(t)+\zeta^k(t)\Bigr)\,rj_2(kr)\,,\nonumber \\
h_k(t,r)&=&\gamma^k(t)\, j_0(kr)\,,\nonumber \\
f_k(t,r)&=&-\frac{3}{2}\Bigl(k\beta_0^k(t)-\fracsm{1}{k}\zeta^k(t)
   \Bigr)\,j_1(kr) \label{solu}\end{eqnarray}
with
\[
\beta_i^k(t)=\beta_i(k)\sin(\omega_i t +\alpha_i),\,i=0,1,2,\quad
\gamma^k(t)=\gamma(k)\sin(\Omega t+\delta),\quad
\zeta^k(t)=c_1 t+c_2,
\]\beq
\omega_0^2=k^2+1,\quad\omega_{1,2}^2=k^2+1\pm 2k\rho,\quad
\Omega^2=k^2+\nu^2,\label{parameters}
\eeq
and $j_i(kr)$ are the spherical Bessel functions. The phase shifts
$\alpha_i$, $\delta$ and the $c_i$ are fixed constants depending on the
initial conditions. The zero mode in \eq{solu} is due to the gauge
freedom; since we have fixed the gauge its amplitude $\zeta^k(t)$ is
zero. The frequencies $\omega_i$, $i=0,1,2$ are eigenmodes of free
gauge bosons (in dimensionful units 1 has to be replaced by $m_W^2$
in \eq{parameters}), $\Omega$ represents the free Higgs particle eigenstates.

The coefficients can be evaluated by Fourier transformation:
\begin{eqnarray}
\gamma^k(t)&=&\fracsm{2k^2}{\pi}\int^\infty_0 dr\,r^2j_0(kr)\,(H(t,r)-1)
\,,\label{fourier} \\
\beta_0^k(t)&=&-\fracsm{4k}{3\pi}\int^\infty_0 dr\,r^2 j_1(kr)\,G(t,r)
   \nonumber \\
&=&\fracsm{2k^2}{9\pi}\int^\infty_0 dr\,r\Bigl[j_0(kr)\,(2B(t,r)+C(t,r))
\nonumber \\
&& \hspace{2cm} {}+j_2(kr)\,(2B(t,r)-2C(t,r))\Bigr]\,,\nonumber \\
\beta_{1,2}^k(t)&=&\fracsm{2k^2}{9\pi}\int^\infty_0 dr\,r
\Bigl[j_0(kr)\,(2B(t,r)+C(t,r))
\nonumber \\
&& \hspace{2cm}{} -j_2(kr)\,(B(t,r)-C(t,r))
\pm 3j_1(kr)\,(A(t,r)-1)\Bigr]\,.\nonumber
\end{eqnarray}

To find the amplitudes $\beta_i(k)$, $\gamma(k)$ we Fourier transform
the numerical solution of the equations of motion \ur{eqnmotion} according
to \eq{fourier}. The resulting functions  $\beta_i^k(t)$, $\gamma^k(t)$
should oscillate according to \eq{parameters} when the system has settled
to small fluctuations about the vacuum. This is the case after some time
$t_{\rm osc}$ (typically $\sim 15\,m_W^{-1}$) with excellent accuracy. 
Moreover we
checked that
the two different formulas for $\beta^k_0(t)$ in \eq{fourier} numerically
lead to the same result. Fitting the functions
$\beta_i^k(t)$, $\gamma^k(t)$ for $t>t_{\rm osc}$ with the corresponding
$\sin(\omega_i t +\alpha_i)$, $\sin(\Omega t +\delta)$
of \eq{parameters} yields the amplitudes $\beta_i(k)$, $\gamma(k)$.

One obtains for the total energy of the linearized system in momentum space
(its coordinate space representation is given by \eq{en2order};
the numerical values coincide up to a deviation of less than $2\%$):
\begin{eqnarray}
E_{\rm tot}^{(2)}&=&E_W+E_H=m_W\int_0^\infty dk\,e_W(k)
   +m_W\int_0^\infty dk\,e_H(k) \nonumber \\
&=&\frac{9\pi^2 m_W}{g^2}\int_0^\infty \frac{dk}{k^2}
   \Bigl(\omega_0^4(k)\beta_0^2(k)+\omega_1^2(k)\beta_1^2(k)
    +\omega_2^2(k)\beta_2^2(k)\Bigr) \nonumber \\
   && {} + \frac{4\pi^2 m_W}{g^2}\int_0^\infty
   \frac{dk}{k^2}\,\Omega^2(k)\gamma^2(k)\,. \label{enmoment}
\end{eqnarray}
Since energy and particle density are related by $e(k)=\omega(k)n(k)$ we
can extract the total number of particles:
\begin{eqnarray}
N_W&=&\frac{9\pi^2}{g^2}\int_0^\infty \frac{dk}{k^2}
   \Bigl(\omega_0^3(k)\beta_0^2(k)+\omega_1(k)\beta_1^2(k)
    +\omega_2(k)\beta_2^2(k)\Bigr)\,, \nonumber \\
N_H&=&\frac{4\pi^2}{g^2}\int_0^\infty
   \frac{dk}{k^2}\,\Omega(k)\gamma^2(k)\,. \label{partnum}
\end{eqnarray}

\section{Results}
\setcounter{equation}{0}

In this section we present the numerical results of our calculation. Our
model contains two free parameters, the Higgs mass $\nu=m_H/m_W$ and the
chemical potential $\rho=\mu/\mu_{\rm crit}$. We performed the calculations
for the values $\nu=0$, 1, 10, and $\rho=-0.2$, $-0.4$, $-0.6$, and $-0.8$.
Moreover we investigated the case $\nu=1$, $\rho=-0.9$. The values for $\nu$
cover a wide range of Higgs masses, but it turns
out that most results do not depend too much on $\nu$. The choice $\nu=0$ is
certainly not physical since essential features of the model, like
spontaneous symmetry breaking, disappear. It should therefore be understood
as limiting case of small masses. In fact we found that the
configurations obtained for $\nu=0$ and a mass like $\nu=0.1$ are
almost identical. Similarly $\nu=10$ is an example for a large
Higgs mass.

\subsection{Barrier penetration}
In Fig.~2 we show the potential and kinetic energy as well as the Chern--Simons
number as functions of the time for the bounce trajectories with chemical
potentials $\rho=-0.2$ and $-0.8$ and the Higgs mass $\nu=1$.
The arbitrariness of the time scale has been removed by setting the time
when the system reaches the escape point at the other side of the barrier
to $t=0$.
It can be seen that the energy conservation law $E_{\rm tot}=T_{\rm kin}
-V_{\rm pot}^\mu\equiv 0$ is excellently fulfilled. For $\rho=-0.8$ the
energies $T_{\rm kin}$ and $V_{\rm pot}^\mu$ are much smaller than for
$\rho=-0.2$. Also the value of $N_{\rm CS}$ at $t=0$ is lower in the case
$\rho=-0.8$. The reason for this behavior is that the barrier between the
trivial vacuum and the topological sector with $N_{\rm CS}=1$ decreases
if $|\rho|$ is increased so that less action is necessary to penetrate
it and the system can escape at a configuration with lower winding number
$N_{\rm CS}$. The extension of the bounce in time (and also in space,
see below), however, increases with $|\rho|$.
For example, in the case $\rho=-0.2$ the system moves out of the trivial
vacuum significantly at about $t\approx -2$ while this happens for $\rho=-0.8$
already at about $t\approx -4$. The Chern--Simons term in the functional $S_E$
(\eq{SEhedge}) lowers the action for large sizes, and its influence 
becomes stronger
with increasing $|\rho|$. Hence for large $|\rho|$ configurations with large
sizes are favored, while for low $|\rho|$ the minimum is taken at
a field configuration with a small size. In the limit $\rho\to 0$ one would
even obtain a configuration with size zero (see below).
\begin{figure} [ht]
\frenchspacing
\centerline{
\epsfxsize=6.in
\epsfbox[85 355 553 700]{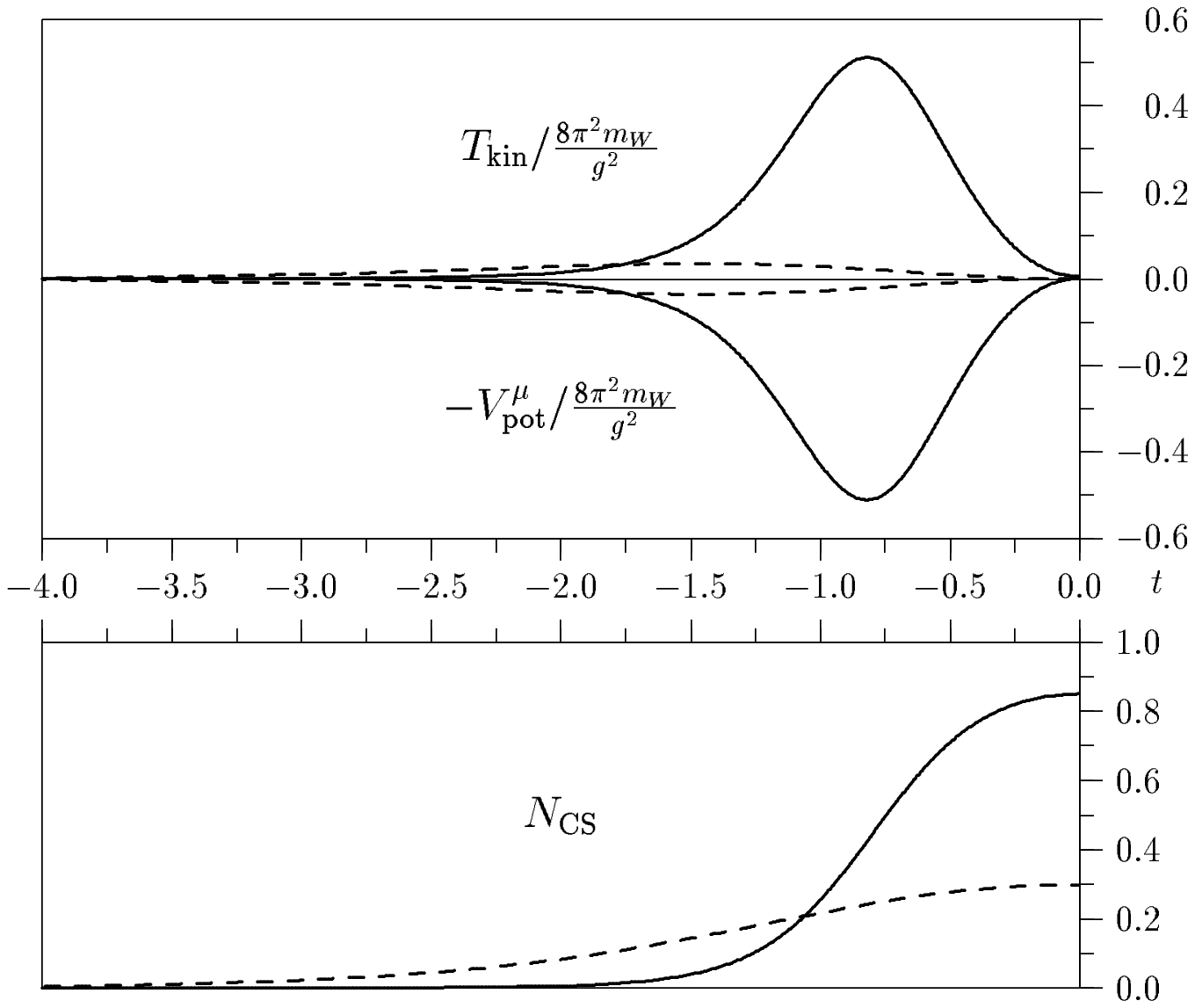}}
\nonfrenchspacing
\begin{quote}\begin{quote}
{\it Fig.~2:\/} The potential energy $-V_{\rm pot}^\mu$, the kinetic
energy $T_{\rm kin}$, and the Chern--Simons number $N_{\rm CS}$
versus time $t$ (in units of $m_W^{-1}$) for chemical potentials
$\rho=\mu/\mu_{\rm crit}=-0.2$ (solid lines) and $-0.8$ (dashed lines).
The Higgs mass is $\nu=m_H/m_W=1$.
\end{quote}\end{quote}
\end{figure}

The field configuration at the escape point of the bounce trajectory
at $t=0$ is interesting by itself, since subsequent calculations like
the investigation of the real time behavior of the system require
only this configuration rather than the complete bounce trajectory.
In Fig.~3 we have plotted the profile functions at $t=0$ again for the
two cases $\rho=-0.2$ and $-0.8$, and the Higgs mass $\nu=1$. What we saw in
Fig.~2 for the time $t$, we find here for the space coordinate $r$:
The deviation of the fields from their values
in the trivial vacuum is much stronger for $\rho=-0.2$, but the region
where they deviate is less extended. A suitable and accurate analytic fit
for the configurations at the escape point is provided in the appendix.
\begin{figure} [ht]
\frenchspacing
\centerline{
\epsfxsize=6.in
\epsfbox[85 345 553 700]{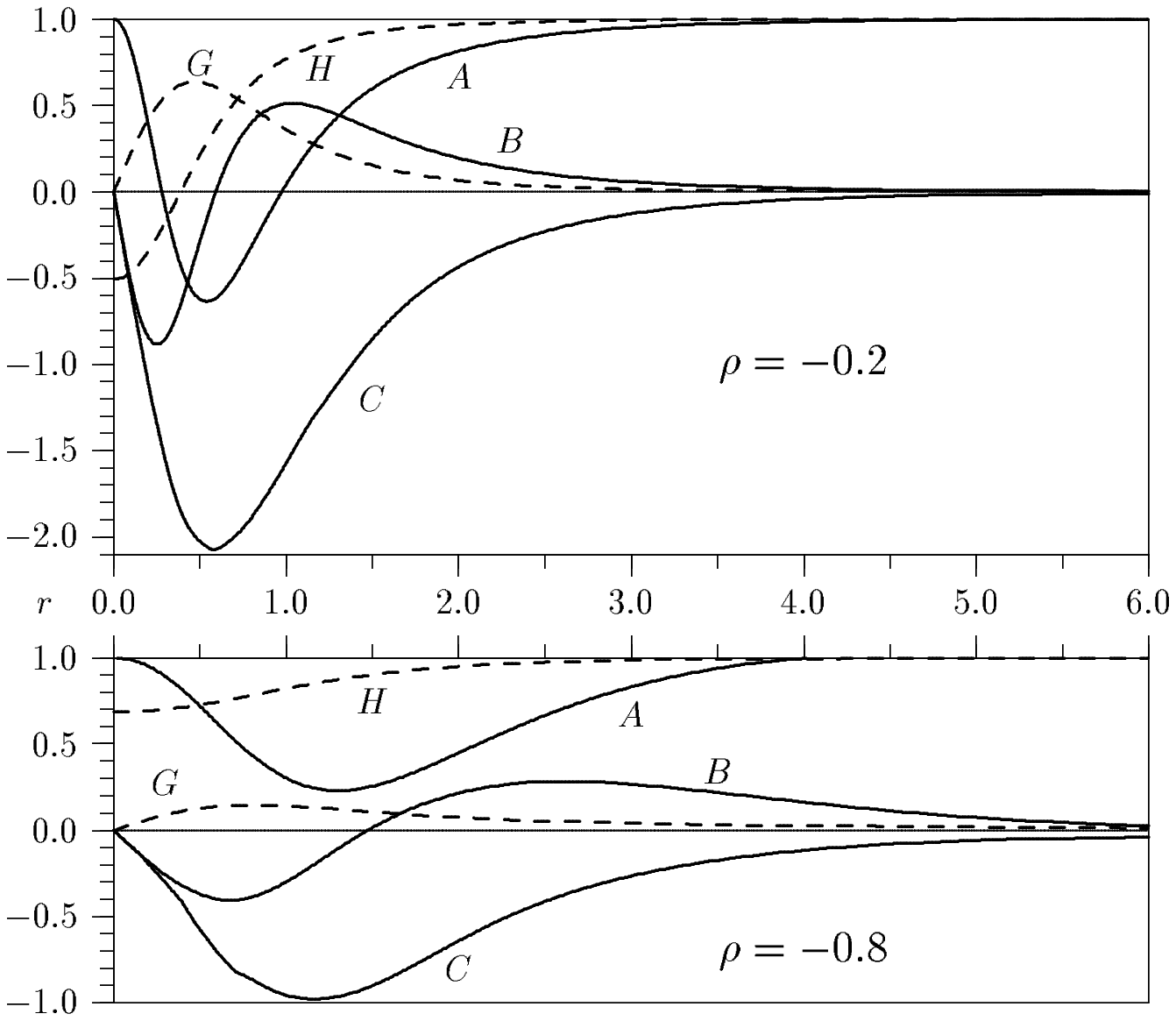}}
\nonfrenchspacing
\begin{quote}\begin{quote}
{\it Fig.~3:\/} The profile functions $A$, $B$, $C$
(solid lines), and $H$ and $G$ (dashed lines) versus the radial distance
$r$ (in units of $m_W^{-1}$) of the configuration at the escape point
of the bounce trajectory for $\rho=\mu/\mu_{\rm crit}=-0.2$
and $\rho=-0.8\,$, and the Higgs mass $\nu=m_H/m_W=1$.
\end{quote}\end{quote}
\end{figure}
\vspace{1.4cm}
\[
\begin{array}{|c||c|c|c|c|c|c|}
\hline
  & \multicolumn{6}{|c|}{\rho=\mu/\mu_{\rm crit}} \\
\hline
\nu=\fracsm{m_H}{m_W} & 0.0 & -0.2 & -0.4 & -0.6 & -0.8 & -1.0 \\
\hline\hline
0  & 1.00^{1)} & 0.81 & 0.54 & 0.30 & 0.11 & 0.00^{1)} \\
\hline
1  & 1.00^{1)} & 0.82 & 0.58 & 0.34 & 0.14 & 0.00^{1)} \\
\hline
10 & 1.00^{1)} & 0.86 & 0.66 & 0.45 & 0.19 & 0.00^{1)} \\
\hline
\multicolumn{7}{l}{\mbox{$^{1)}$ known from theory}} \\
\end{array}
\]
\begin{quote}
{\it Tab.~1:\/}
The action $S_E/S_{\rm inst}$ of the bounce trajectory for various values
of the chemical potential $\rho$ and the Higgs mass $\nu$.
\end{quote}
\vspace{0.7cm}
\[
\begin{array}{|c||c|c|c|c|c|c|}
\hline
  & \multicolumn{6}{|c|}{\rho=\mu/\mu_{\rm crit}} \\
\hline
\nu=\fracsm{m_H}{m_W} & 0.0 & -0.2 & -0.4 & -0.6 & -0.8 & -1.0 \\
\hline\hline
0  & 1.00^{1)} & 0.83/0.85^{2)} & 0.64/0.66^{2)}
&  0.44/0.47^{2)} & 0.22/0.24^{2)} & 0.00^{1)} \\
\hline
1  & 1.00^{1)} & 0.85 & 0.68 & 0.51 & 0.30 & 0.00^{1)} \\
\hline
10 & 1.00^{1)} & 0.87 & 0.74 & 0.66 & 0.42/0.42^{2)} & 0.00^{1)} \\
\hline
\multicolumn{7}{l}{\mbox{$^{1)}$ known from theory}} \\
\multicolumn{7}{l}
{\mbox{$^{2)}$ results of two different trajectories with the same action}} \\
\end{array}
\]
\begin{quote}
{\it Tab.~2:\/}
The Chern--Simons number $N_{\rm CS}^{\rm esc}$ of the escape point
of the bounce trajectory for various values
of the chemical potential $\rho$ and the Higgs mass $\nu$.
\end{quote}

\noindent
In Table 1 we give the results for the action $S_E$ of the tunneling
process in units of the action $S_{\rm inst}=\frac{8\pi^2}{g^2}$ of the
instanton in pure gauge theory. In Table 2 we show
the Chern--Simons number $N_{\rm CS}^{\rm esc}$ of the
configuration at the escape point. For $\rho=-1$ the barrier vanishes so
that the tunneling process is reduced to a single point in the
configuration space, namely the trivial vacuum. Hence in this case the
action and $N_{\rm CS}^{\rm esc}$ are both $0$. For $\rho\to 0$ the field
configuration which minimizes the action tends to an instanton
with size zero, so that
the bounce action equals the instanton action and the configuration at the
escape point is a vacuum with $N_{\rm CS}=1$. In a pure gauge field
theory the solution for $\rho=0$ would be an instanton of arbitrary
size, but here the scale invariance is destroyed by the non-zero
vacuum expectation value of the Higgs field, so that for a finite
size the action would be larger than the instanton action \cite{Affleck1}.
Hence in the limit $\rho\to 0$ we obtain a trajectory with size zero.
For $-1<\rho<0$, however, due to the Chern--Simons term the minimum
of the action is taken by a configuration with finite size.
The accuracy of the data in Tables 1 and 2 can be estimated by
increasing the density of the lattice (which is usually of size about
$81\times 81$) and the number of sweeps (usually of the order of 10000).
We find that the numerical error of the results is around $1\%$.
For each set of parameters $\nu$, $\rho$ we performed two independent
minimizations, starting from two rather different field configurations
like e.g.~instantons with size $\lambda=2$ and $\lambda=4$ (see \eq{iniconf}).
For $\nu=1$, both minimizations always ran towards the same bounce
trajectory, within the given frame of accuracy.
For $\nu=0$ and $\nu=10$, however, the two minimizations sometimes
produced different trajectories,
which have the same action, but different $N_{\rm CS}^{\rm esc}$, and
also different behavior of the functions $V_{\rm pot}^\mu(t)$,
$N_{\rm CS}(t)$. Using a suitable averaging procedure, one finds that
there exists an infinite number of different paths, which
all have the same action (up to a deviation of $1\%$). Two conclusions
are possible: Either the bounce trajectory is not unique, i.e the
action has a zero mode, or there is a unique, but very
shallow minimum.

Unfortunately, the numerical accuracy of our method does not allow to
distinguish between those two possibilities, but for the following
investigations this is quite irrelevant, anyway: In the case of a true
zero mode the different tunneling trajectories will be taken with
{\it exactly} the same probability while in the case of a shallow minimum
the tunneling probabilities are {\it almost} the same. We will see below
that the real time evolution of the fields after the tunneling leads
to deviating bosonic signatures for two different trajectories. In
any case, if a barrier penetration happens, both results occur
(almost) equally
likely, so that the accuracy of all results is given by the range
of values which we obtain for different trajectories with
the same action. Certainly, the tunneling rate is influenced by the volume
of a possible transformation group with invariant action or the (low)
curvature around the minimum, but these only contribute to the prefactor
$B$ of \eq{rate} which is not discussed in this paper.

Thus, in Tab.~2 we have given both values for $N_{\rm CS}^{\rm esc}$
if we obtained two different configurations with the same action.
We find that both the action and $N_{\rm CS}^{\rm esc}$ increase
if $|\rho|$ is decreased or the Higgs mass $\nu$ is increased. The
reason is that the barrier becomes wider and higher with decreasing
$|\rho|$ and increasing $\nu$ so that more action is necessary to
penetrate through it, and the escape point moves further away from
the trivial vacuum. We see that both quantities are roughly linearly related
to $\rho$, and, as anticipated, the dependence of the results on the
Higgs mass is rather weak.

The suppression factor $e^{-2S_{\rm inst}}\approx 10^{-153}$ of the
tunneling rate for $\rho=0$ becomes less strong for $|\rho|>0$,
but significant tunneling amplitudes
can only be obtained for chemical potentials as high as
$|\rho|\simgt 0.9$ (for $\rho=-0.9$ and $\nu=1$ we obtained
$S_E/S_{\rm inst}=0.06$ and $N_{\rm CS}^{\rm esc}=0.17$).
One needs, however, a matter density which is about $10^6$ times larger than
the photon density in the early universe at the electroweak phase
transition, or about $10^{18}$ nuclear matter density, in order to correspond 
to such a large chemical potential.
Presently it is not known if a matter density
of this order has ever existed in the early universe.
Even if this is not the case, our results still have some physical
significance because the tunneling rate may be related to the rate
of fermion number violation at high particle energies \cite{DiPet}.
Hence, taking our results one might be able to deduce the probability
to observe such a process at a supercollider.

\subsection{Real time evolution}
Next we will describe how the system evolved in real time after the
barrier penetration. We solved the equations of motion basically as it was
done e.g.~in \cite{Hellmund,Cott}. In order to check the numerics,
we first used the sphaleron as starting configuration, and our results
agreed with those of \cite{Hellmund,Cott}. Then we replaced the sphaleron
by the configuration at the escape point of the barrier penetration
process. By performing the gauge transformation with
$\Delta N_{\rm CS}=-1$ to ensure that the fields fluctuate about the
{\it trivial} vacuum (see Section 4)
the potential energy of the starting configuration $V_{\rm pot}^\mu$
is increased from 0 to
$|\mu|$ (in units of $\frac{8\pi^2}{g^2}m_W=\frac{\mu_{\rm crit}}{2}$
this means it is increased from 0 to $2|\rho|$). The Chern--Simons number
is lowered by one unit and starts between $-1$ and $0$.

In Fig.~4 we present the behavior of $V_{\rm pot}^\mu$, $T_{\rm kin}$,
and $N_{\rm CS}$ as a function of time $t>0$ after the tunneling
process for $\rho=-0.6$ and $\rho=-0.9$. The Higgs mass is $\nu=1$.
We find that for times $t\simlt 10$ the behavior of the system in
the two cases is quite similar: The system starts to move, i.e.~the
kinetic energy increases, while the potential energy decreases.
Energy conservation is fulfilled very accurately during the whole
process. The Chern--Simons number increases quickly to values around
0, which means the systems comes close to the trivial vacuum.
\begin{figure} [ht]
\frenchspacing
\centerline{
\epsfxsize=6.in
\epsfbox[85 335 553 700]{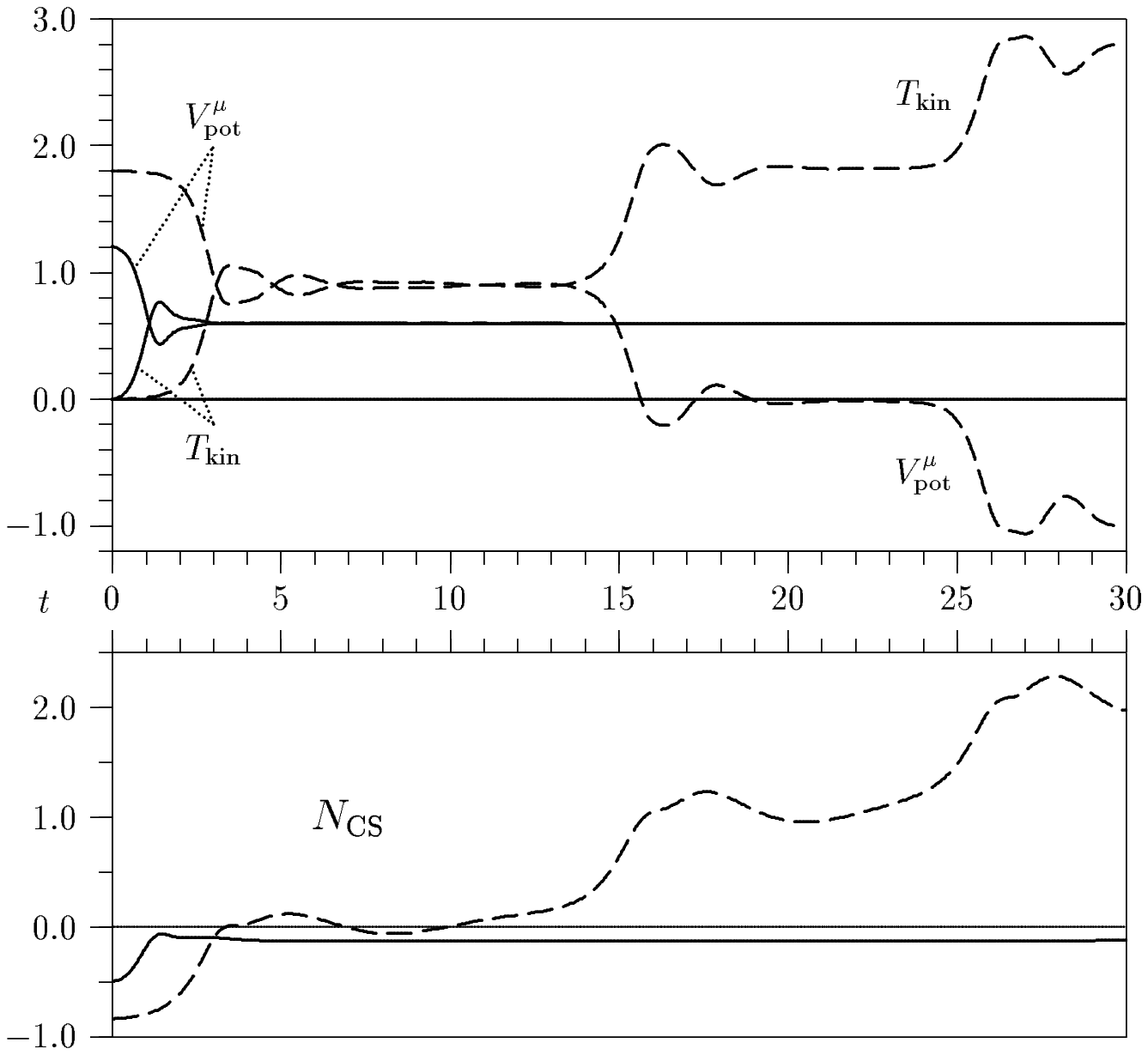}}
\nonfrenchspacing
\begin{quote}\begin{quote}
{\it Fig.~4:\/} The potential energy $V_{\rm pot}^\mu$, the kinetic energy
$T_{\rm kin}$ (in units of $\frac{8\pi^2 m_W}{g^2}$), and the Chern--Simons
number $N_{\rm CS}$ versus time $t$ (in units of $m_W^{-1}$) for chemical
potentials $\rho=\mu/\mu_{\rm crit}=-0.6$ (solid lines) and $-0.9$
(dashed lines), and the Higgs mass $\nu=m_H/m_W=1$.
\end{quote}\end{quote}
\end{figure}
 
In the case $\rho=-0.6$ the energy dissipates into small fluctuations about
the vacuum. We checked that the energy ${\Vpotm}^{(2)}$ in second order
of the fluctuations agrees up to a deviation of less than $1\%$ with
$V_{\rm pot}^\mu$, moreover we see in Fig.~4 that potential and
kinetic energy both become constant at the value of $\frac{|\mu|}{2}$ in
accordance with the virial theorem. The Chern--Simons number
takes a constant value slightly below zero. Hence the system has settled
to small oscillations about the trivial vacuum and will stay in this
topological sector forever (apart from possible tunneling later on).
Later we will analyze the particle content of this state.

In the case $\rho=-0.9$, however, we observe a completely different
behavior for times $t\simgt 10$. Here $N_{\rm CS}$ suddenly increases
from values around 0 to about 1, later even to 2. Hence, the system does
not stay in the topological sector of the trivial vacuum but moves
classically over the next barrier to the sector with $N_{\rm CS}=1$.
Here it also stays only for a short period before it moves to the
next sector with $N_{\rm CS}=2$. This behavior is also demonstrated
by the plot of the potential energy which shows the successive falls of
the system like a cascade towards configurations with increasing winding
number and decreasing energy. Once the first tunneling process
has happened, the system moves classically over all the following barriers
so that the whole fermion matter
decays rapidly and sets free an enormous amount of energy. As mentioned above,
the tunneling amplitude is not extremely small for
$\rho=-0.9$ ($10^{-9}$ instead of $10^{-153}$), but in order to generate 
such a large chemical potential a huge fermion density is required
($\approx 10^{18}$ nuclear matter density).

It is a property of the periodic plus linear potential that even at
small $\mu$, the energy barriers become lower
than the local minimum at $\Ncs=0$, if one goes far enough in
$\Ncs$. Therefore, if the systems tunnels directly to that far-away
sector (which would require multi-instanton-like bounce solutions),
the avalanche would probably develop, too. Of course, the
multi-instanton tunneling probability is even smaller than for a single bounce,
but it should grow faster with $\mu$. It would be interesting to
estimate the total decay probability as a function of $\mu$, with
tunneling to different topological sectors summed up.

Energetically, this behavior of a classical rapid decay is allowed if
the top of the barrier between the sectors with $N_{\rm CS}=0$ and 1 is
lower than the chemical potential $|\mu|$. For $\nu=1$, this is the case
already for $|\rho|\simgt 0.2$, but we found that only for
$|\rho|\simgt 0.9$ it actually happens. For chemical potentials between
0.2 and 0.9 the system could in principle cross the next barrier, but
the energy is dissipated among the modes of small oscillations and
not concentrated on the direction to the next minimum so that the system
does not find the collective path over the barrier. For $\nu=0$, we
have found that the avalanche starts developing
already at $|\rho|\simgt 0.8$.
We think we have observed an interesting phenomenon of how an
exponentially suppressed spontaneous decay triggers off a catastrophic
avalanche which never stops until the fermion sea, originally filled
up to the Fermi surface $\mu$, is completely ``splashed''.

In Fig.~5 we show for $\nu=1$ and $\rho=-0.6$ how the density of the
total energy, defined by
$E_{\rm tot}=\Tkin+\Vpotm\equiv m_W\int_0^\infty dr\,e_{\rm tot}(r)$,
evolves in time. Our plot is similar to the one given in
\cite{Hellmund,Cott}, where one starts with a slightly disturbed
sphaleron instead of the configuration at the escape point of the
bounce. As was found in \cite{Hellmund,Cott}, the outgoing wave moves
with almost the speed of light and shows some dispersion, but
in our case the dispersion is less strong. For the sphaleron,
after $t=25\,m_W^{-1}$ the height of the
pulse has decreased to about $30\%$ of its original value at $t=0$,
while for the bounce configuration it drops only to about $65\%$ of the
value at $t=0$.
\begin{figure} [ht]
\frenchspacing
\centerline{
\epsfxsize=6.in
\epsfbox[85 435 553 700]{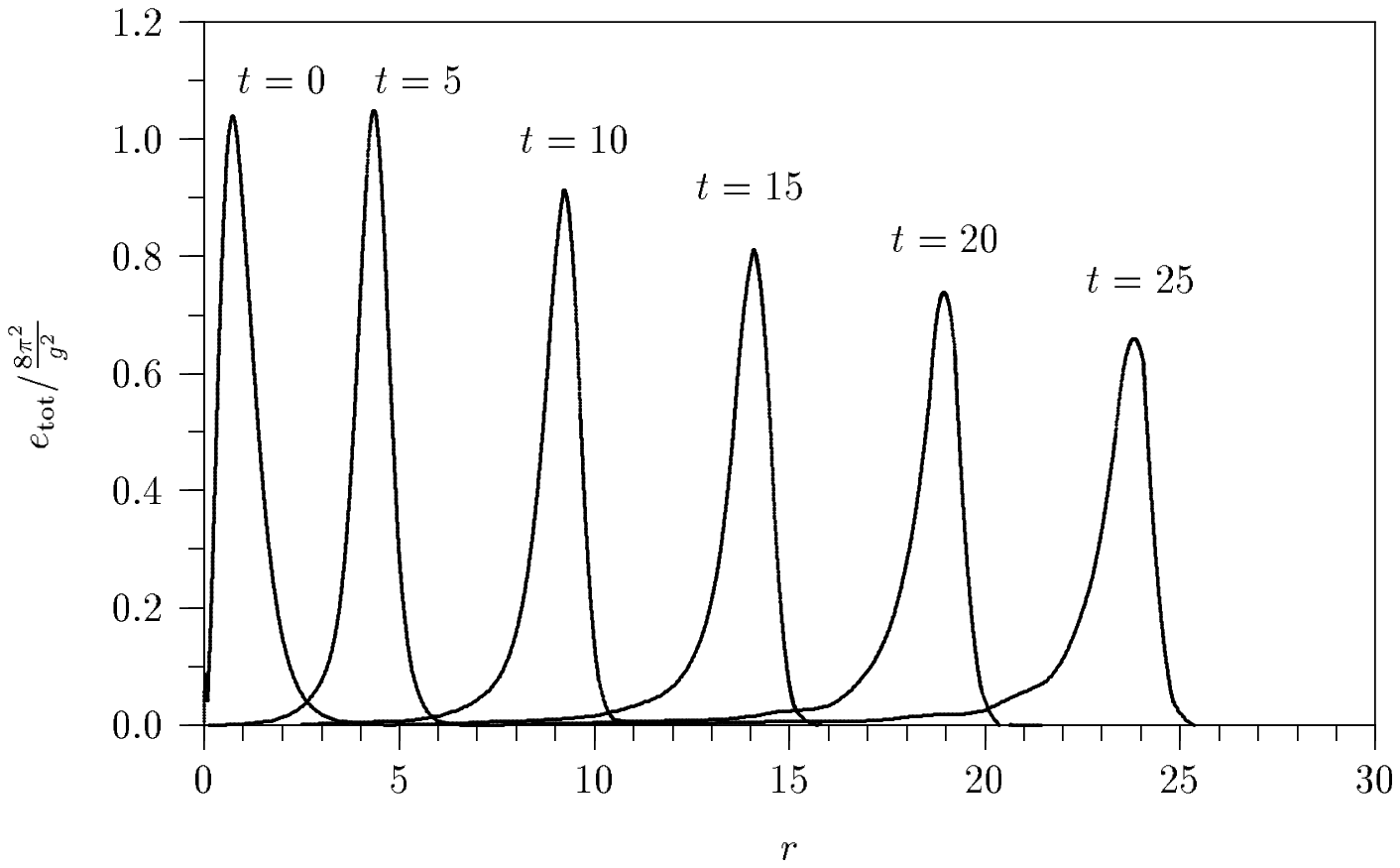}}
\nonfrenchspacing
\begin{quote}\begin{quote}
{\it Fig.~5:\/} The density $e_{\rm tot}$ of the total energy versus radial
distance $r$ for various times $t$ ($r$ and $t$ in units of $m_W^{-1}$).
The parameters are $\rho=\mu/\mu_{\rm crit}=-0.6$ and $\nu=m_H/m_W=1$.
\end{quote}\end{quote}
\end{figure}

Next we wish to analyze the particle content of the state after the
tunneling. This is only possible if the system stays in the topological
sector of the trivial vacuum and does not move classically over the next
barrier. In this case after some time (typically $\sim 10\,m_W^{-1}$)
the system has settled to small oscillations about the trivial vacuum
so that one can perform the Fourier decomposition \ur{fourier}.
Fig.~6 shows how the total energy ($=|\mu|$) is distributed among the
Higgs and gauge boson modes. Integration of the curves
yields the total energy of the Higgs ($E_H$) and gauge ($E_W$) bosons
(see \eq{enmoment}); we
have found that the sum $E_H+E_W$ is equal to $|\mu|$ up to a deviation of
usually less than $2\%$, which is another check of our numerics. 
\vspace{0.7cm}
\[
\begin{array}{|c||c|c|c|c|}
\hline
  & \multicolumn{4}{|c|}{\rho=\mu/\mu_{\rm crit}} \\
\hline
\nu=\fracsm{m_H}{m_W} &\phantom{--} -0.2\phantom{--}
& \phantom{--} -0.4 \phantom{--} & \phantom{--} -0.6 \phantom{--}
& \phantom{--} -0.8 \phantom{--} \\
\hline\hline
0  & 3.8/3.1^{1)} & 4.9/4.2^{1)} & 6.4/6.3^{1)} & --^{2)} \\
\hline
1  & 5.1 & 7.6 & 8.5 & 11.3 \\
\hline
10  & 0.0 & 0.0 & 2.3 & 1.3/1.5^{1)} \\
\hline
\multicolumn{5}{l}
{\mbox{$^{1)}$ results of two different trajectories with the same action}}
   \\
\multicolumn{5}{l}
{\mbox{$^{2)}$ system does not oscillate about trivial vacuum}} \\
\multicolumn{5}{l}{\mbox{\phantom{$^{2)}$ }but moves classically to next
   sector}} \\
\end{array}
\]
\begin{quote}
{\it Tab.~3:\/}
The ratio of the energy of the Higgs bosons to the total energy
$E_H/(E_W+E_H)$ in percent after the system settles to small oscillations
about the trivial vacuum. The results are given for various values
of the chemical potential $\rho$ and the Higgs mass $\nu$.
\end{quote}
 
In Tab.~3
we show how much of the energy is taken by the Higgs bosons (in percent).
This number is generally in the range up to $10\%$; it increases slightly with
$|\rho|$.
If we increase the Higgs mass from $\nu=0$ to $\nu=1$ (and keep $\rho$ fixed),
the Higgs particles gain some energy on the expense of the gauge
bosons, but for $\nu=10$ the share of the Higgs bosons is almost zero.
In this last case the Higgs bosons are too heavy to be produced at all,
for small masses their total energy is basically correlated to the individual
energy of each particle, i.e.~it rises with the mass.

Fig.~6 demonstrates that the
spectrum is shifted to larger $k$ when $\nu$ is increased. This effect is
particularly strong in the case of the Higgs bosons.
For $\nu=10$ the energy
density takes its maximum at about $k\approx 5$ while for $\nu\simlt 1$
it is only at $k\approx 1.5$, for $\nu=0$ very light Higgs bosons with
momenta around $k=0$ are produced in a large number.
\begin{figure} [ht]
\frenchspacing
\centerline{
\epsfxsize=6.in
\epsfbox[85 385 553 700]{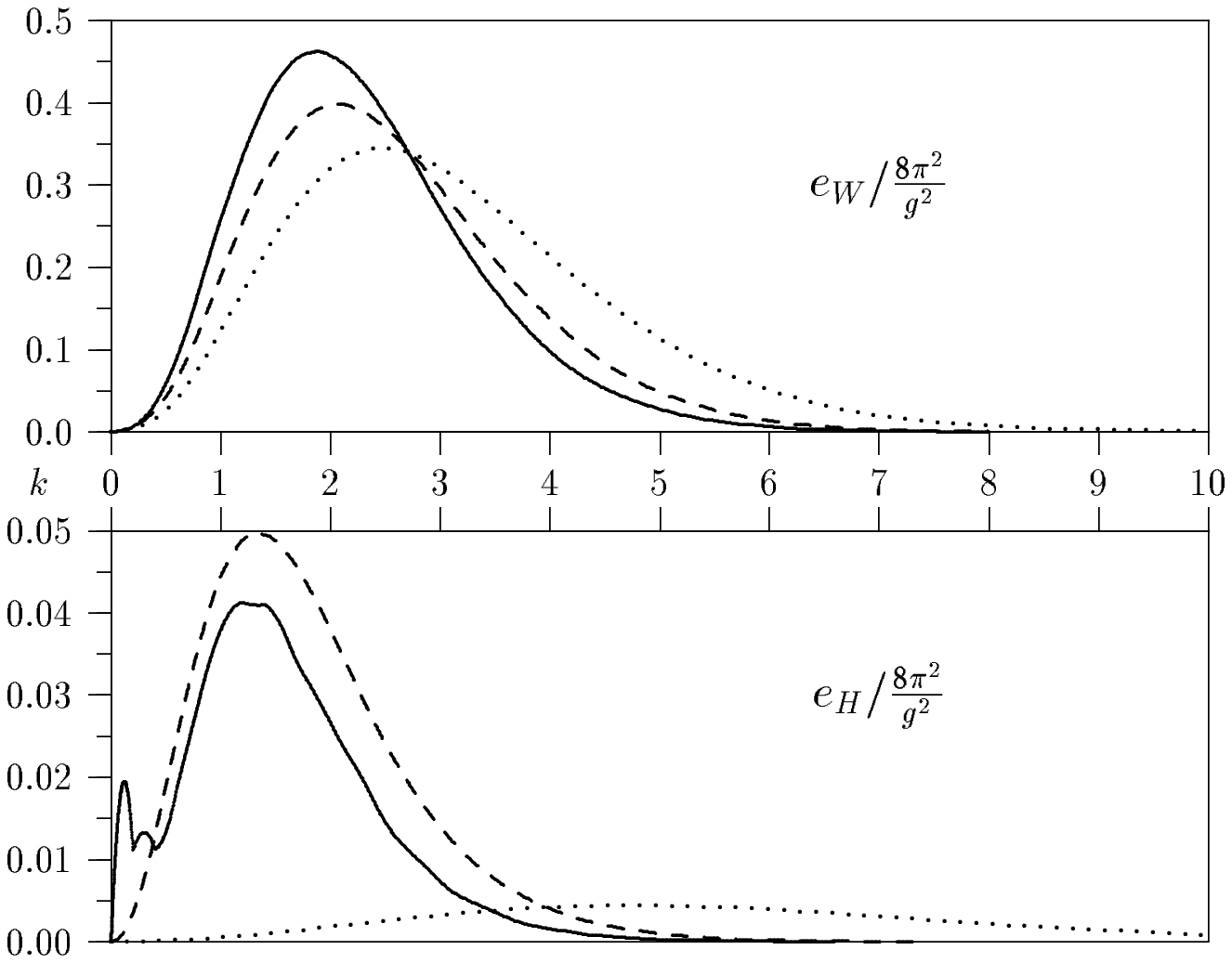}}
\nonfrenchspacing
\begin{quote}\begin{quote}
{\it Fig.~6:\/} The energy densities $e_W$ and $e_H$ of the gauge and
Higgs bosons versus momentum $k$ (in units of $m_W$) 
for Higgs masses $\nu=m_H/m_W=0$
(solid lines), $\nu=1$ (dashed lines), and $\nu=10$ (dotted lines).
The chemical potential is $\rho=-0.6$.
\end{quote}\end{quote}
\vspace{-0.5cm}
\end{figure}

Tab.~4 shows the total particle numbers $N_W$ and $N_H$ of the gauge and Higgs
bosons, respectively. In the case $\nu=0$ the determination of $N_H$ is not
possible because the number density $n_H(k)$ is strongly peaked close to $k=0$
so that the numerical error of the integration is uncontrollable. This is of
course an artefact of the unphysical choice $\nu=0$, however, also for finite, 
but small masses one would have to take many values in the $k$-lattice
around 0 and perform the integration carefully to get a reasonable result. 
We find that the
particle numbers rise with $|\rho|$, because the released energy increases
and allows the production of more particles. The number of Higgs particles
is much smaller than the number of gauge bosons, again we see that in case
of a large Higgs mass no Higgs bosons are produced.
\vspace{0.7cm}
\[
\begin{array}{|c||c|c|c|c|}
\hline
  & \multicolumn{4}{|c|}{\rho=\mu/\mu_{\rm crit}} \\
\hline
\nu=\fracsm{m_H}{m_W} \rule[-3mm]{0mm}{8mm} & -0.2 & -0.4 & -0.6 & -0.8 \\
\hline\hline
0\rule[-5mm]{0mm}{13mm}  & \atopla{24.5/22.7^{1)}}{--^{2)}}
& \atopla{49.0/45.5^{1)}}{--^{2)}}
& \atopla{74.2/69.3^{1)}}{--^{2)}} & \atopla{--^{3)}}{--^{3)}} \\
\hline
1 \rule[-5mm]{0mm}{13mm}  & \atopla{22.9}{1.9} & \atopla{46.2}{5.9}
& \atopla{67.1}{9.6} & \atopla{102.6}{16.8} \\
\hline
10 \rule[-5mm]{0mm}{13mm}   & \atopla{20.8}{0.0} & \atopla{40.3}{0.0}
& \atopla{64.3}{0.4} & \atopla{117.2/112.4^{1)}}{0.3/0.4^{1)}} \\
\hline
\multicolumn{5}{l}
{\mbox{$^{1)}$ results of two different trajectories with the same action}}
   \\
\multicolumn{5}{l}
{\mbox{$^{2)}$ determination impossible due to infrared behavior}} \\
\multicolumn{5}{l}
{\mbox{$^{3)}$ system does not oscillate about trivial vacuum}} \\
\multicolumn{5}{l}{\mbox{\phantom{$3)$ }but moves classically to next
   sector}} \\
\end{array}
\]
\begin{quote}
{\it Tab.~4:\/}
The number of gauge bosons $N_W$ (upper numbers) and Higgs bosons
$N_H$ (lower numbers) after the system settles to small oscillations
about the trivial vacuum. The results are given for various values
of the chemical potential $\rho$ and the Higgs mass $\nu$.
\end{quote}

\noindent
Finally we comment on the numerical uncertainty of the data in Tabs.~3 and 4.
The error can be estimated by increasing the numerical parameters of the
Runge--Kutta time integration and the number of $k$ values in the Fourier
transformation. Moreover one can choose different times $t_{\rm osc}$
(see explanation after \eq{fourier})
where we start to fit the amplitudes of the modes in momentum space.
We find that the error of the data is in general less than $2\%$.
We have to keep in mind, however, that the bounce trajectory, and hence
the starting configuration, is not unique, but in some cases there are
different solutions with almost the same tunneling probability. These different
starting configurations yield results for the particle content which can
deviate up to $20\%$, as can be seen from the data in Tabs.~3 and 4.
Therefore, the probability density for the particle content of the final
state is spread over a
range of numbers about $\pm 10\%$ around the value given in
the tables.

\section{Summary}
\setcounter{equation}{0}

In this work we have presented a method to find the bounce trajectory
in the electroweak theory and calculated the probability for the decay of
high density fermionic matter.

The bounce trajectory is obtained by minimization of the Euclidean action
as a function of the discretized Higgs and gauge boson fields.
At each step of the procedure the action is regarded as a function of
only one parameter, i.e.~it is minimized with respect to the value of one
field profile function at a certain point in the lattice while the values
of the other fields and at the other points are kept fixed. After finishing one
step of the minimization one moves to the next field or next point until each 
field at each point of the lattice
has been considered. Many of such ``sweeps'' through the lattice (of the
order of 10000) have to be performed until a stable configuration is reached
which does not change any more if it undergoes further sweeps. From time to
time the user has to interfere into the process of minimization. The program
contains several options to manipulate the field configuration, partially
in order to keep the fields in a continuous and smooth shape, partially in
order to accelerate the convergence. It has been
checked that the final configuration always fulfills the Euclidean equations
of motion with sufficient accuracy.

The determination of the bounce has been carried out for several choices of
the Higgs mass $\nu=m_H/m_W$ and the chemical potential
$\rho=\mu/\mu_{\rm crit}$ of the fermionic matter. We find that the 
action $S_E$
of the bounce drops from the instanton action $S_{\rm inst}$ at $\rho=0$ to
zero at $|\rho|=1$ roughly linearly and depends only weakly on $\nu$. A similar
behavior is found for the Chern--Simons number of the escape point of the
bounce, $N_{\rm CS}^{\rm esc}$, which decreases from 1 to 0.
The action $S_E$ is the exponent of the tunneling rate which itself is
correlated to the probability of the fermion number violation at high
particle energies. It might therefore be possible to use our results for
$S_E$ in order to predict the cross section of the high energy process.

For several sets of parameters we found that the bounce solution is 
not unique;
instead there exist several solutions with different escape points,
but with the same action (in the given frame of accuracy).
Since after the barrier penetration each of
these escape points will be taken with the same probability, for some
results of this work we can only give a range of values
instead of a definite number.

After the tunneling process, the bosonic fields
can evolve in real time Min\-kow\-skian
space since they have obtained the energy of the annihilated fermions as
potential energy. The equations of motion can here be solved by some
time integration method rather than a minimization of the action.

For chemical potentials $|\rho|\simlt 0.8$ the system
stays in the topological sector where it came to after the tunneling
and settles to small oscillations about the minimum.
The oscillations correspond to the radiation of the
Higgs and gauge bosons, and we have analyzed
the particle content of this state by Fourier transformation. We find
that usually less than $10\%$ of the energy is absorbed by the creation of 
the Higgs bosons, and correspondingly the total number of produced 
gauge bosons is also about 10 times greater than the number of Higgs
bosons. The results depend strongly on $\rho$ and partially also on
the Higgs mass $\nu$. For large chemical potentials  $|\rho|\simgt 0.9$
the system has enough energy and coherence after the tunneling
to move classically over the next barriers. This corresponds to
an avalanche decay
of the fermionic matter and to the production of an enormous amount of
Higgs and gauge bosons.

\medskip
{\bf Acknowledgements:}
We are grateful to C.Weiss for pointing our attention to the use of
relaxation methods (\cite{Adler}), and thank P.Pobylitsa, M.Polyakov,
and V.Petrov
for numerous discussions. The work has been supported in part by the Deutsche
Forschungsgemeinschaft and the RFBR grant 95-07-03662.

\setcounter{section}{0}
\def\thesection{Appendix}
\def\theequation{A.\arabic{equation}}
\newpage

\section{}
\setcounter{equation}{0}

To investigate the real time behavior after tunneling
one only needs the field configuration at the escape point of
the bounce (which we fix at $t=0$) rather than the functions $A(t,r),\,\ldots,
\,G(t,r)$ in the whole two-dimensional space $t,r$. Hence it is useful to
have a parameterization of the functions $A(0,r),\,\ldots,\,G(0,r)$ so that
one can take them as input for further calculations without having
the necessity
to recalculate the complete bounce trajectory. In this appendix we give
an analytic fit which matches the numerically determined functions
$A(0,r),\,\ldots,\,G(0,r)$ very accurately. The potential energy and the
Chern--Simons number of the fit agree with the corresponding values
of the numerical configuration up to a deviation of about $1\%$. Following the
real time behavior of the fit and the numerical fields we find that
the particle numbers $N_W$, $N_H$ and the energies $E_W$, $E_H$ are
reproduced up to a deviation less than $3\%$.

The parameterization is chosen so
that it includes the possibility to describe both the trivial vacuum
with $N_{\rm CS}=0$ and the non-trivial vacuum with $N_{\rm CS}=1$.
For this reason the fit is performed in a gauge where the field
$D(r)\equiv 0$ everywhere. We denote the other fields in this gauge by
$A_0(r)$, $B_0(r)$, $H_0(r)$, $G_0(r)$. (Here and in the following the
argument $t=0$ is dropped.) The gauge transformation
which transforms $A_0(r),\,\ldots,\,G_0(r)$ to the field configuration
$A(r),\,\ldots,\,G(r)$ at the escape point of the bounce is described by
some function $P(r)$ according to \eq{hedgau}.

For the functions $P(r)$, $D(r)$, $A_0(r)$, $B_0(r)$, $H_0(r)$,
$G_0(r)$ we use the following ansatz:
\begin{eqnarray}
P(r)&=&-\lambda_d\biggl[d_0\Bigl(1+\fracsm{r}{2\lambda_d}\Bigr)
+d_2\Bigl(1+\fracsm{r}{\lambda_d}+\fracsm{r^2}{2\lambda_d^2}\Bigr) \nonumber
   \\
&& \hspace{3cm}{}
   +3d_3\Bigl(1+\fracsm{r}{\lambda_d}+\fracsm{r^2}{2\lambda_d^2}
+\fracsm{r^3}{6\lambda_d^3}\Bigr)\biggr] e^{-r/\lambda_d}\,, \nonumber \\
D(r)&=&2P'(r)=\biggl[d_0\Bigl(1+\fracsm{r}{\lambda_d}\Bigr)
+d_2\fracsm{r^2}{\lambda_d^2}+d_3\fracsm{r^3}{\lambda_d^3}\biggr]
e^{-r/\lambda_d}\,, \nonumber \\
A_0(r)&=&a_0\biggl(1+\fracsm{r}{\lambda_a}+a_2\fracsm{r^2}{\lambda_a^2}
+a_3\fracsm{r^3}{\lambda_a^3}\biggr) e^{-r/\lambda_a} + 1 \,, \nonumber \\
B_0(r)&=&b_0\biggl(1+\fracsm{r}{\lambda_b}+b_2\fracsm{r^2}{\lambda_b^2}
+b_3\fracsm{r^3}{\lambda_b^3}\biggr) e^{-r/\lambda_b} \,, \nonumber \\
H_0(r)&=&\biggl[h_0\Bigl(1+\fracsm{r}{\lambda_h}\Bigr)+h_1 r
+h_2\fracsm{r^2}{\lambda_h^2}
+h_3\fracsm{r^3}{\lambda_h^3}\biggr] e^{-r/\lambda_h} + 1 \,, \nonumber \\
G_0(r)&=&\biggl[g_0\Bigl(1+\fracsm{r}{\lambda_g}\Bigr)+g_1 r
+g_2\fracsm{r^2}{\lambda_g^2}+g_3\fracsm{r^3}{\lambda_g^3}
+g_4\fracsm{r^4}{\lambda_g^4}\biggr] e^{-r/\lambda_g} \,.\label{fitfunctions}
\end{eqnarray}
The procedure how we obtain the parameters in \eq{fitfunctions} is 
the following:
To ensure the correct behavior of the fitting functions at $r=0$ (\eq{origin})
we first take
$d_0=B'(0)$ (evaluated by a quadratic fit of $B$ at $r=0$), use a suitable
fitting algorithm for $D(r)$ to determine $\lambda_d$, $d_2$, and $d_3$, 
and set
\begin{eqnarray}
a_0&=&\cos \Bigl(2\lambda_d(d_0+d_2+3d_3)\Bigr) -1\,, \nonumber \\
b_0&=&\sin \Bigl(2\lambda_d(d_0+d_2+3d_3)\Bigr)\,,  \nonumber \\
h_0&=&H(0) \cos\Bigl(\lambda_d(d_0+d_2+3d_3)\Bigr) -1\,, \nonumber \\
g_0&=&H(0) \sin\Bigl(\lambda_d(d_0+d_2+3d_3)\Bigr)\,,  \nonumber \\
h_1&=&\Bigl(H(0)\fracsm{d_0}{2}-G'(0)\Bigr)
\sin\Bigl(\lambda_d(d_0+d_2+3d_3)\Bigr)\,, \nonumber \\
g_1&=& -\Bigl(H(0)\fracsm{d_0}{2}-G'(0)\Bigr)
\cos\Bigl(\lambda_d(d_0+d_2+3d_3)\Bigr)\,, \label{fitparam}
\end{eqnarray}
where we have used $P(0)=-\lambda_d(d_0+d_2+3d_3)$. The parameters
$\lambda_a$, $\lambda_b$, $\lambda_h$, $\lambda_g$, are held fixed:
\beq
\lambda_a=\lambda_b=0.8\,,\hspace{2cm}\lambda_h=\lambda_g=0.6\,.
\label{fitlambda}\eeq
Then we perform a gauge transformation on the fields
$A(r),\,\ldots ,\,G(r)$, using the
function $P(r)$ of \eq{fitfunctions} to obtain $A_0(r),\,\ldots ,\,G_0(r)$.
These functions are fitted to yield the
remaining parameters $a_2$, $a_3$, $b_2$, $b_3$, $h_2$, $h_3$,
$g_2$, $g_3$, and $g_4$. Altogether, our fits contain 12 parameters
determined by the fitting algorithm plus 7 parameters depending on
the three values $B'(0)$, $H(0)$, and $G'(0)$. So in total the number
of free parameters is 15.

In Table 5 the results of the parameters
are given for several values of the chemical potential $\rho$ and
the fixed Higgs mass $\nu=m_H/m_W=1$. For this Higgs mass we always
obtain a unique field configuration at the escape point, i.e. it
does not depend on the initial configuration before the minimization
of the action starts.
\vspace{0.7cm}
\[
\begin{array}{|c||c|c|c|c|c|}
\hline
\rho=\mu/\mu_{\rm crit} & -0.2 & -0.4 & -0.6 & -0.8 & -0.9 \\
\hline
\hline
a_0 & -1.166 & -1.873 & -1.957 & -1.379 & -0.685 \\
\hline
a_2 & -0.123 & -0.183 & -0.108 & +0.358 & +1.100 \\
\hline
a_3 & -0.042 & -0.006 & -0.020 & -0.084 & -0.144 \\
\hline
\hline
b_0 & +0.986 & +0.487 & -0.290 & -0.925 & -0.949 \\
\hline
b_2 & -0.738 & -0.472 & -1.008 & -0.138 & +0.326 \\
\hline
b_3 & +0.180 & +0.270 & -0.303 & -0.224 & -0.349 \\
\hline
\hline
\lambda_d & +0.258 & +0.320 & +0.343 & +0.528 & +0.705 \\
\hline
d_0 & -5.514 & -3.165 & -2.224 & -1.001 & -0.501 \\
\hline
d_2 & -3.298 & -2.524 & -1.926 & -1.341 & -0.740 \\
\hline
d_3 & +0.001 & -0.005 & +0.000 & +0.162 & +0.118 \\
\hline
\hline
h_0 & -0.674 & -0.976 & -0.957 & -0.625 & -0.299 \\
\hline
h_1 & +0.663 & +0.974 & +0.963 & +0.538 & +0.228 \\
\hline
h_2 & +0.153 & +0.129 & +0.118 & +0.031 & -0.011 \\
\hline
h_3 & -0.025 & -0.016 & -0.015 & -0.005 & +0.000 \\
\hline
\hline
g_0 & +0.389 & +0.092 & -0.285 & -0.561 & -0.506 \\
\hline
g_1 & -0.556 & -0.254 & +0.145 & +0.360 & +0.315 \\
\hline
g_2 & -0.141 & -0.104 & -0.043 & +0.009 & +0.011 \\
\hline
g_3 & +0.134 & +0.117 & +0.066 & -0.027 & -0.057 \\
\hline
g_4 & -0.019 & -0.013 & -0.004 & +0.010 & +0.018 \\
\hline
\end{array}
\]
\begin{quote}
{\it Tab.~5:\/}
The parameters of the fitting functions for the configuration at the
escape point of the bounce trajectory according to \eq{fitfunctions}
for several
values of the chemical potential $\rho$ and the Higgs mass $\nu=m_H/m_W=1$.
\end{quote}

\end{document}